\documentclass[master,       
               twoside,        
               BCOR=10mm,       
              bibliography=totoc,
               paper=a4,
               pagesize,
               ngerman,english,  
               final,          
               ]{GAUBM}

\usepackage[firstpage]{draftwatermark}
\usepackage{ifluatex}
\ifluatex
\thisShouldGenerateAnError!!
    \usepackage[utf8]{luainputenc}
\else
    \usepackage[]{fontenc}
    \usepackage[utf8]{inputenc}
\fi
\usepackage{amsmath,amssymb} 
\usepackage{mleftright}
\usepackage[type1]{libertine}
\usepackage[libertine,bigdelims,vvarbb,liby]{newtxmath}

\usepackage[scaled=0.92]{inconsolata}
\usepackage{fix-cm}

\usepackage{collcell} 

\usepackage{setspace}  
\AtBeginDocument{%
  \addtolength\abovedisplayskip{-0.5\baselineskip}%
  \addtolength\belowdisplayskip{-0.5\baselineskip}%
}
\usepackage{csquotes} 
\usepackage[ngerman,english]{babel}     
\usepackage{calc}      

\usepackage{scrhack}

\usepackage{pdflscape} 
\usepackage[figuresright]{rotating}

\usepackage{multirow}
\usepackage{booktabs}                      
\usepackage{longtable}                     
\usepackage{array}                         

\usepackage{textcomp,gensymb}

\usepackage[version=3]{mhchem}
\usepackage{enumitem} 

\usepackage{pdfpages}

\def\ensurenomath#1{\ifmmode \mbox{#1}\else#1\fi} 
\usepackage{xspace} 
\usepackage{textcomp} 
\usepackage{xfrac}

\usepackage{xcolor}
\usepackage{placeins} 
\usepackage{fancyvrb, listings} 

\usepackage[final]{microtype}
\usepackage[
per-mode=fraction,
separate-uncertainty,
retain-unity-mantissa=false,
table-alignment=center,
detect-all, 
binary-units=true,
exponent-product=\cdot,
exponent-base=10
]{siunitx}
\DeclareSIUnit[number-unit-product = {}]\AU{a.u.}
\DeclareSIUnit[number-unit-product = {}]\pixel{pixel}
\usepackage{cancel}

\usepackage[
textsize=footnotesize,
obeyFinal,
]{todonotes}

\usepackage{icomma} 

\usepackage[sorting=none,natbib,citestyle=numeric-comp,backend=biber,
backref=true,
firstinits=true,alldates=iso8601]{biblatex}
\DeclareSourcemap{
	\maps[datatype=bibtex]{
		\map[overwrite=true]{
			\step[fieldset=abstract, null]
		}
	}
}
\DeclareDocumentCommand{\posscite}{ o o m }{\citeauthor{#3}'\IfNoValueTF{#1}{s}{#1}\IfNoValueF{#2}{ #2} \cite{#3}}

\usepackage{mathtools}

\DeclarePairedDelimiter\floor{\lfloor}{\rfloor}
\DeclarePairedDelimiter\abs{\lvert}{\rvert}

\usepackage{xstring}
\newcommand*{\repr}[1]{%
    \IfEqCase*{#1}{%
    {all}{all}%
    {sino}{sinogram-selected}%
    {TB}{Grimm et al.}%
    {integr}{sinogram-selected}%
    {man}{manually-selected}%
    }[\todo{Did not match any case!!}]%
}%

\usepackage{url}

\usepackage{pgffor}

\usepackage{wallpaper}

\usepackage{subcaption}
\usepackage[
margin=10pt,
labelformat=simple,
]{caption} 

\usepackage{bm}

\newcommand*{\data}[1]{\textsc{#1}\xspace}
\newcommand*{\dd}{
  \mathop{}\!\mathrm{d}%
}

\DeclareMathOperator*{\mean}{mean}

\newcommand*{\defd}{\mathop{:=}}

\newcommand*{\vect}[1]{\bm{#1}}
\newcommand*{\norm}[1]{\left\lVert#1\right\rVert}

\newcommand*{\matlab}{\texttt{MATLAB}\xspace}
\newcommand*{\nlinvpp}{\texttt{nlinv++}\xspace}

\newcommand*{\me}{\mathrm{e}}
\newcommand*{\mi}{\mathrm{i}}
\providecommand{\uppi}{\pi}

\DeclareDocumentCommand\rf{ m g }{%
{\ang{#1}-pulse\IfNoValueF{#2}{#2}\xspace}}

\usepackage{layout}

\usepackage[
sanitizesort=true,
]
{glossaries}
\setacronymstyle{long-short}
\makeindex
\makeglossaries
\newacronym{CT}{CT}{computed tomography}
\newacronym{DFT}{DFT}{discrete Fourier transform}
\newacronym{FID}{FID}{free induction decay}
\newacronym{FLASH}{FLASH}{fast low-angle shot}
\newacronym{FOV}{FOV}{field of view}
\newacronym{GPU}{GPU}{graphics processing unit}
\newacronym{GRAPPA}{GRAPPA}{generalized autocalibrating partially
parallel acquisition}
\newacronym{IRGNM}{IRGNM}{iteratively regularized Gauss-Newton method}
\newacronym{MRI}{MRI}{magnetic resonance imaging}
\newacronym{NLINV}{NLINV}{non-linear inverse reconstruction}
\newacronym{NMR}{NMR}{nuclear magnetic resonance}
\newacronym{PCA}{PCA}{principal component analysis}
\newacronym{PSF}{PSF}{point spread function}
\newacronym{rf}{rf}{radio frequency}
\newacronym{SENSE}{SENSE}{sensitivity encoding}
\newacronym{SNR}{SNR}{signal-to-noise ratio}
\newacronym{SVD}{SVD}{singular value decomposition}
\newacronym[sort={T1}]{T1}{\ensuremath{T_1}}{spin-lattice relaxation time}
\newacronym[sort={T2}]{T2}{\ensuremath{T_2}}{spin-spin relaxation time}
\newacronym[sort={T2s}]{T2s}{\ensuremath{T_2^*}}{effective spin-spin relaxation time}

\usepackage[pdfstartview=FitH,      
            breaklinks=true,        
	bookmarks=false
            ]{hyperref}

\usepackage[
noabbrev,
capitalise,
]{cleveref}

\typearea[current]{current}
\begin{document}
\ThesisAuthor{Hans Christian Martin}{Holme}
\PlaceOfBirth{Soest}
\ThesisTitle{Vorbereitende Datenanalyse zur Rekonstruktion von Echtzeit-MRT-Daten}%
{Preparatory data analysis for the reconstruction of real-time MRI data}
\FirstReferee{Prof.\ Dr.\ Jens Frahm}
\Institute{Biomedizinische NMR Forschungs GmbH\\am Max-Planck-Institut für biophysikalische Chemie}
\SecondReferee{Prof.\ Dr.\ Tim Salditt}
\ThesisBegin{18}{02}{2016}
\ThesisEnd{01}{01}{1970}
\frontmatter
\ThisCenterWallPaper{0.14}{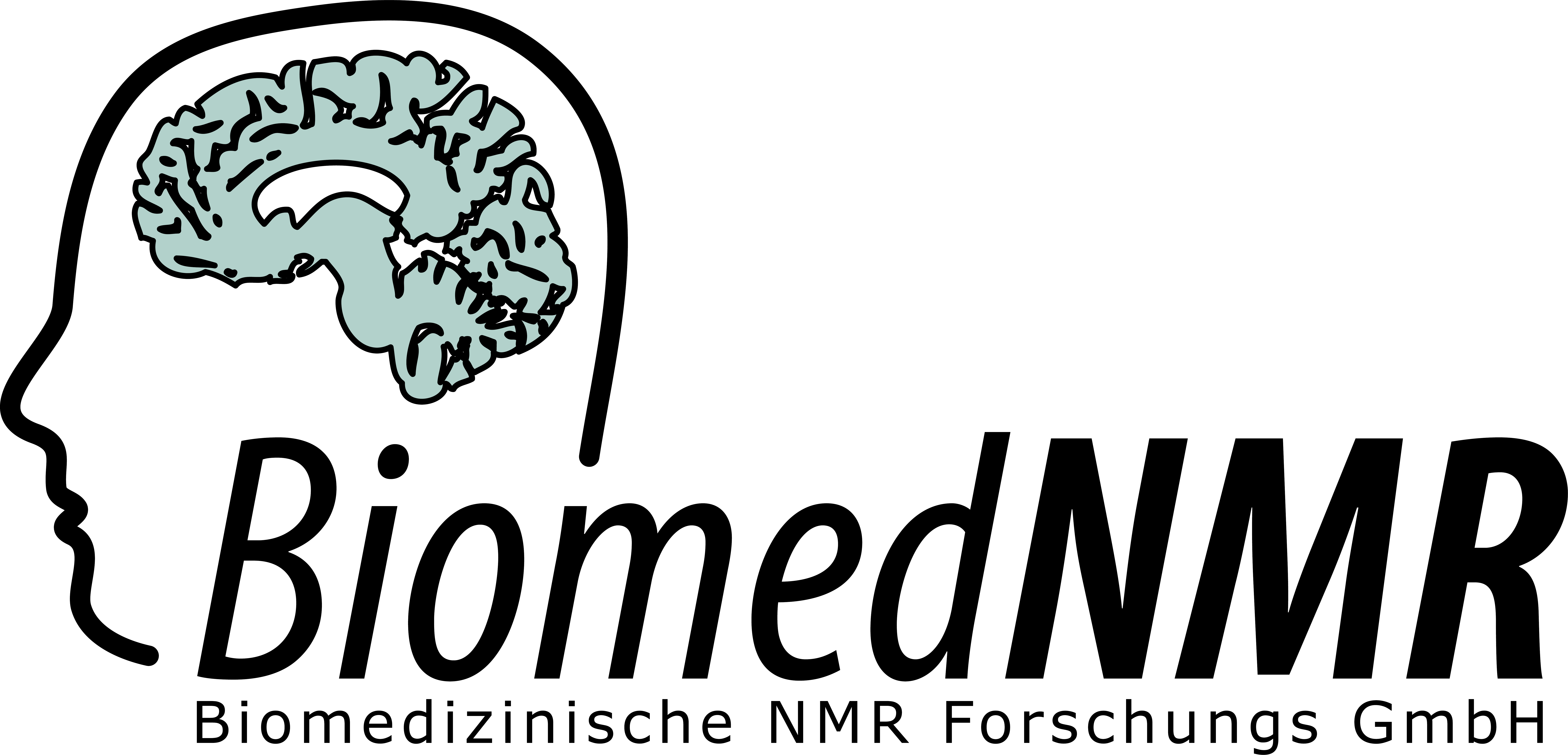}
\setlength{\wpYoffset}{-200pt}

\maketitle

\cleardoublepage


\makeatletter
\@openrightfalse
\tableofcontents
\@openrighttrue
\makeatother

\printglossary[
style=listdotted
]



\listoftodos

\mainmatter   

\chapter{Introduction}
\label{sec:Introduction}
\Gls{MRI} is an important radiological technique for imaging
sections or volumes in clinical and scientific settings. Starting from
its introduction by \textcite{lauterbur} in \citeyear{lauterbur}, it has
evolved into a routine procedure. Because of the lack of ionizing radiation,
its excellent soft tissue contrast, and its wide variety of imaging
modalities, it is now used in hospitals and research institutes worldwide.

An inherent problem of \gls{MRI} is its low speed, since images are
sampled line-by-line in Fourier space. This low inherent speed has
practical implications, e.g. movement during the acquisitions
degrades image quality, and dynamical processes impose requirements
on image timing. Therefore, acceleration of \gls{MRI} image acquisition
has been a research focus since its beginning.

A success in this drive for speed has been the introduction of
\gls{FLASH} imaging in \citeyear{FLASHpatent}\cite{FLASHpatent,FLASH1,FLASH2},
which reduced the imaging
time from minutes to mere seconds. This hastened the integration
of \gls{MRI} into routine clinical use. Further speed improvements
based on \gls{FLASH} \gls{MRI} were possible by tackling the sampling:
The introduction of multiple receiver coils, combined with
parallel \gls{MRI} \cite{parallelMRI}, allows image reconstruction
even from undersampled data. Pushing this undersampling further leads
to the common trade-off of increased imaging speed at the cost of
decreased \gls{SNR}.

Nowadays, accelerated acquisition sequences coupled with
sophisticated reconstruction algorithms to cope with the
high undersampling
have pushed this even further,
with imaging times below \SI{30}{\milli\second} \cite{rt2}. This enables
the recording of \gls{MRI} movies instead of still images.
Together with modern computer hardware, this led to the introduction of
real-time \gls{MRI}, with both image acquisition and image
reconstruction in real time.

While holding great promise, real-time \gls{MRI} also poses new
challenges: The increased imaging speeds, for example,
lead to large volumes of data which, together with
computationally intensive reconstruction algorithms, can diminish
reconstruction speeds. Furthermore, the high degree of undersampling
can lead to unacceptable image artifacts.
This thesis introduces and
evaluates approaches for handling these two problems. \Cref{sec:Compression}
discusses coil compression methods for reducing the data volume, while
a new coil selection algorithm for streak artifact
reduction is introduced in \cref{sec:Selection}.
\enlargethispage{\baselineskip}

\FloatBarrier
\cleardoublepage
\chapter{Magnetic Resonance and Image Reconstruction}
\label{sec:theo}
This chapter provides a brief overview over \gls{NMR} and \gls{MRI},
and also an introduction into the reconstruction technique used
in this work. A more general introduction into \gls{MRI}
and the underlying physics is given in
textbooks like \textcite{handbook} and \textcite{haacke}.

\section{Magnetic Resonance Imaging}
\label{sec:MR}
\subsection{Principles}
\label{sec:MR_principles}
This section follows Chapters 1 and 9 of \textcite{haacke}.

\Gls{MRI} relies on the interaction of a magnetic moment $\mu$ with an
external field $\vect{B_0}$. The magnetic moment in this case is
the nuclear magnetic moment, generated by the nuclear spin.
The principal nucleus used in \gls{MRI} is hydrogen in water, fat,
and other organic molecules.

For a hydrogen nucleus --- a proton --- in a magnetic field $\vect{B_0}$,
the Zeeman effect leads to an energy difference between parallel and
anti-parallel alignment of the spin, with alignment parallel to
$\vect{B_0}$ having
lower energy. The energy difference is $\hbar\gamma B_0$, where
$\hbar$ is the reduced Planck constant and $\gamma$ is the
gyromagnetic ratio, a nucleus-dependent value.

While this constrains the nuclear magnetic moment in the direction of
$\vect{B_0}$ to two possible values, there is no such restriction in the
transversal plane orthogonal to $\vect{B_0}$. Therefore, in analogy to a
classical
magnetic moment, the expectation value will start to precess around
$\vect{B_0}$. The frequency
of this precession is called the Larmor frequency $\omega_0$ with
$\omega_0 = \gamma B_0$. In an extended object made up of
multiple spins, since there is no phase coherence imposed on
this precession, the transversal magnetic moments cancel out, leaving
a bulk magnetization parallel to $\vect{B_0}$.

Since the energy difference between alignment parallel and anti-parallel
to $\vect{B_0}$ is
small compared to the thermal energy at body temperature,
the population difference between the
two states is also small --- at \SI{1}{\tesla}, the population
difference is about 6 excess spins per million
\citep[p. 36]{siemens_introduction} ---
but it still leads to a macroscopic magnetization
$M_0$ which is large
enough for \gls{NMR} effects. $M_0$ is given by \citep[p. 5]{haacke}
$M_0 = \frac{\rho_0\gamma^2\hbar^2}{4kT}B_0$,
where $\rho_0$ is the number of protons per unit volume, $k$ is
the Boltzmann constant, and $T$ is the temperature.

Precession of the macroscopic magnetization can be achieved by
resonant excitation. The resonance condition
is that the frequency of the magnetic excitation pulse $\vect{B_1}(t)$
needs to be the Larmor frequency. For the commonly used nuclei and field
strengths this
frequency is in the radio-frequency range, the excitation pulses
are therefore commonly called \glsreset{rf}\gls{rf} pulses. The angle by
which the magnetization is tipped is called the flip angle, and it depends
both on the field strength and duration of the \gls{rf} pulse.
The excitation pulse achieves its effect by two mechanisms:
First, it will flip spins from
the lower energy parallel to the higher energy anti-parallel state,
thereby reducing
longitudinal magnetization.
Second, it will impose its phase onto the precession, thereby causing a
non-vanishing transversal magnetization. Combined, this leads to a
bulk magnetization
that starts to precess and to tip over the duration of the \gls{rf} pulse.

Suppose an \gls{rf} pulse is applied that tips the magnetization
into the transversal plane (called a \rf{90}).
Directly after this pulse, the bulk magnetization precesses in
the transversal plane. This magnetization
will induce a changing magnetic flux in nearby coils, which can be
detected.

The longitudinal magnetization will eventually relax back into
the equilibrium state.
This can be understood as the spins
excited into the anti-parallel state flipping back into the
parallel state.
This is an exponential process characterized by the
\glsreset{T1}\gls{T1}. If the equilibrium magnetization is $M_0$, the
longitudinal magnetization will recover according to:
\begin{align*}
  M_\|(t) = M_0\left(1-\me^{-\frac{t}{T_1}}\right).
\end{align*}
Furthermore, the different spins are not only affected by the
external magnetic field, but also by the magnetic fields produced by
other spins. This leads to slightly different Larmor frequencies for
each spin, causing a loss of phase coherence called dephasing
and thereby a loss of bulk transversal magnetization. This decay is
characterized by the \glsreset{T2}\gls{T2}, following
\begin{align*}
 M_\perp(t) = M_0\me^{-\frac{t}{T_2}}.
\end{align*}
\gls{T2} is always shorter than \gls{T1}, with typical values for
\gls{T2} on the order of \SIrange{1}{100}{\milli\second} and for
\gls{T1} on the order of \SIrange{100}{1000}{\milli\second}.
Inhomogeneities of the static
magnetic field $\vect{B_0}$ will lead to even faster dephasing,
characterized by the \glsreset{T2s}\gls*{T2s} with $\gls{T2s} < \gls{T2}$.

\begin{figure}[htbp]
\centering
\begin{subfigure}{0.45\textwidth}
  \centering
 \includegraphics[width=0.90\textwidth]
 {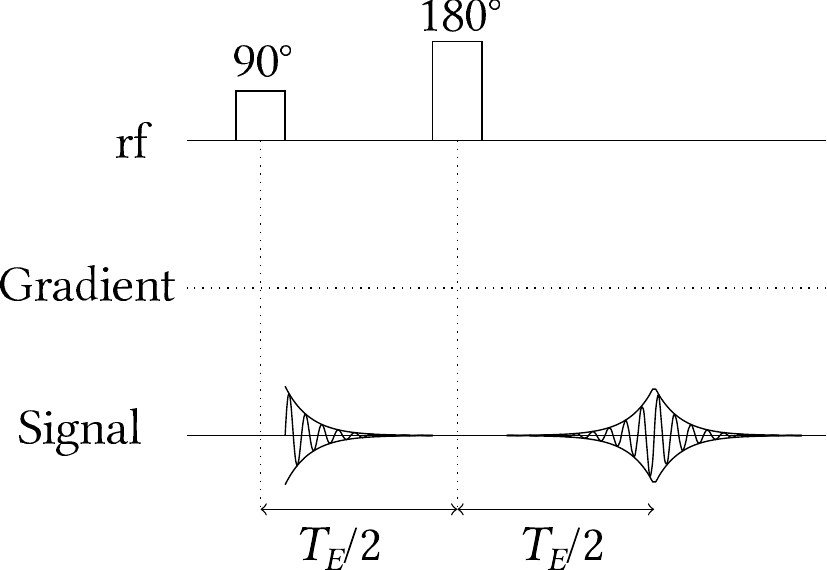}
 \caption{}
 \label{fig:SE}
\end{subfigure}
\hspace{0.05\textwidth}
\begin{subfigure}{0.45\textwidth}
  \centering
 \includegraphics[width=0.90\textwidth]
 {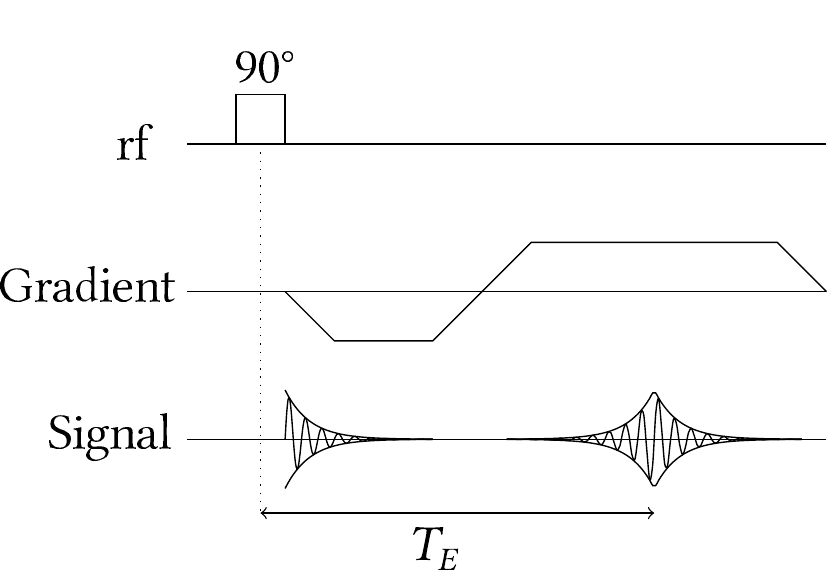}
 \caption{}
 \label{fig:GRE}
\end{subfigure}
\caption[]{
Diagram for \subref{fig:SE} spin echo and \subref{fig:GRE} gradient echo
formation. For the spin echo in \subref{fig:SE}, no gradients are necessary:
after an inital \ang{90} \gls{rf}-pulse,
the \gls{FID} signal begins to decay. Applying a \rf{180} after
$\sfrac{T_E}{2}$ leads to a spin echo after $T_E$. The gradient echo
in \subref{fig:GRE} shows the same \gls{FID} decay after the \rf{90},
but the spins are then dephased and rephased with an applied gradient.
Plotted here is the gradient strength over time.
Diagrams adapted from \cite{mueckerphd}.
}
\label{fig:echoes}
\end{figure}

Application of an excitation pulse to an object generates
a decaying signal called \glsreset{FID}\gls{FID}.
An alternative way to create a signal is by generating an echo.
The basic setup is the same: a \rf{90} is applied to tip the magnetization
into the transversal plane, where they will dephase due to \gls{T2} decay.
By applying a second \gls{rf}-pulse after $\tau$, this time  a
\rf{180}, the magnetization stays in the transversal
plane but it is rotated by \ang{180} about an axis within the transversal
plane; or alternatively, the phase of each spin is turned to its
negative. This means that the spins that acquired excess phase between
excitation and \rf{180} now lag behind and vice versa. Therefore,
the spins rephase at $t=2\tau$ and an echo forms.
This is called a spin echo. The strength is still attenuated by
$\me^{-\sfrac{2\tau}{T_2}}$ due to \gls{T2} decay. \Cref{fig:SE} shows
a diagram of spin echo formation.

An echo can also be generated without using a second \gls{rf} pulse.
For this, the spins are first dephased by applying a gradient and then
rephased by applying a gradient with the same strength but with the
opposite sign. The echo is formed at the instant where the
combined area under
both gradients equals zero. The resulting echo is called
a gradient echo or
gradient-recalled echo. In contrast to a spin echo, the
gradient echo is attenuated by the shorter \gls{T2s}, since the
magnetic field inhomogeneities are not canceled out by phase inversion.
\Cref{fig:GRE} depicts gradient echo formation.

For generating an image, it is necessary to relate spatial position
and signal. The idea in \gls{MRI} is to use magnetic gradient fields
and exploit the linear dependence of the Larmor frequency on field
strength. This can be understood most easily in the one-dimensional case:
By applying a linear gradient $G(t)$ in
$x$-direction the total magnetic field becomes $B(x,t) = B_0 + xG(t)$.
This changes the local Larmor frequency to
$\omega(x,t) = \omega_0 + \Delta\omega(x,t)$ with
$\Delta\omega(x,t) = \gamma xG(t)$. This is called frequency encoding,
since the position of the spins is hereby
related to their precession frequency. The signal $s(t)$ in
\gls{MRI} depends on
the effective spin density\footnote{The effective spin density is the
plain spin density $\rho_0$ with
all constants absorbed into it. This simplifies the resulting expressions.
See p. 141 of \textcite{haacke} for an exact definition.}
$\rho(\vect{r})$ and the phase of the precessing
spins $\phi(\vect{r},t)$ \cite[p. 141]{haacke},
\begin{align*}
 s(t) = \int\rho(\vect{r})\me^{\mi(\omega_0 t + \phi(\vect{r},t))}\dd^3r
\end{align*}
where $\phi(\vect{r},t) = -\int_0^t\omega(\vect{r}, t)\dd t$.
For a one-dimensional object $\rho(\vect{r}) = \rho(x)\delta(y)\delta(z)$,
the signal after frequency encoding and demodulation of $\omega_0$
becomes
\begin{align*}
 s(t) = \int\rho(x)\me^{-\mi\int_0^t\Delta\omega(x,t)\dd t}\dd x.
\end{align*}
By noting that $\int_0^t\Delta\omega(x,t)\dd t = \gamma z\int_0^tG(t)\dd t$
and introducing $2\uppi k(t) \defd \gamma\int_0^tG(t)\dd t$, the time
dependence can be made implicit in $k$:
\begin{align*}
 s(k) = \int\rho(x)\me^{-2\uppi \mi kx}\dd x.
\end{align*}
Here, $s(k)$ is the Fourier transform of $\rho(x)$, and so can be inverted
to obtain the spin density from a frequency-encoded measurement
 $\rho(x) = \int s(k)\me^{2\uppi \mi kx}\dd x$.
The $k$ is naturally identified
as a vector in the reciprocal or Fourier space, called $k$-space in
\gls{MRI}. The temporal evolution of the gradient determines $k(t)$, which
is referred to as the trajectory in $k$-space. In practice, the signal
cannot be recorded continuously but is sampled at discrete points instead.

The gradient echo
introduced above is an example of a trajectory that can be used to sample
a region around the center of $k$-space: The first dephasing gradient
moves the sampling point away from the $k$-space center
for a time $\tau$. If the subsequent
rephasing gradient has the same gradient strength but twice the duration,
a symmetric line around the $k$-space center will be encoded in the signal.
If it is sampled during the rephasing gradient, this symmetric
$k$-space line can be used to reconstruct the object by 1D Fourier
transformation.

For two-dimensional objects, there are several encoding schemes.
One is radial encoding, which will be explained in \cref{sec:radial_FLASH},
another is Cartesian encoding. For Cartesian encoding, parallel lines are
sampled in $k$-space. Each line of $k$-space is measured with
frequency encoding in a separate repetition, so with a separate excitation
pulse.
To measure shifted lines in $k$-space an additional gradient,
called phase-encoding gradient, is applied
between excitation and frequency encoding. The direction of this gradient
has to be orthogonal to the frequency-encoding direction. This gradient
leads to a phase shift in real space, which in turn implies a translation
in $k$-space.
For a certain choice of
line separation, gradient strength and sampling rate, the measured samples
will lie on a Cartesian grid in $k$-space, hence the name Cartesian
encoding.
The reconstruction is simply a two-dimensional discrete Fourier transform.
\Cref{fig:Cartesian} shows a diagram of a Cartesian encoding scheme.

For imaging a slice of a three-dimensional object, slice selection can be
used. Here, a gradient orthogonal to the desired slice is applied during
the excitation pulse. If the gradient is such that it is zero in the
desired slice, the Larmor frequency changes everywhere
except in the slice itself, so the resonance condition is only fulfilled
in the slice and only it will be excited, yielding effectively a
two-dimensional object for imaging.

For three-dimensional imaging, either phase encoding can be extended in
the third direction, or slice selection can be used to image each slice
serially.


Image reconstruction in \gls{MRI} invariably involves discretization, which
introduces the problem of sampling: the imaging procedures described
above can only work if proper sampling is taken into account, with
the sampling rate subject to the Nyquist criterion and the
maximum $k$-space extent determining the resolution \citep[Ch. 12]{haacke}.
Sampling and associated artifacts are discussed in more detail in
\cref{sec:artifacts}.

In general, the choice and timing of imaging gradients and \gls{rf} pulses
can modify the contrast of \gls{MRI} images to an astonishing degree,
with far more possibilities than could be described here. An
overview over a large number of \gls{MRI} sequences can be found in
\textcite{handbook}.

Modern \gls{MRI} uses multiple receive coils. An advantage is the
increased \glsreset{SNR}\gls{SNR}, which comes from smaller coils which
can be placed closer to the object. Multiple coils with overlapping
sensitivity profiles can cover the same volume as a single large coil.
But, instead of the single global image reconstructed above, each coil
contains information from only part of the object. Therefore, a combined
image is normally calculated as the root-sum-of-squares of the individual
coil images \cite{RSS}.
A further advantage of multiple receive coils is the possibility of
parallel imaging:
2D and especially 3D images require a large number of repetitions, leading
to long measurement times.
The idea of parallel imaging is to speedup this process by
using the spatial information contained in the sensitivity variation
of the coils to replace time-consuming spatial encoding.

In Cartesian imaging, the speedup comes
from skipping phase-encoding lines.
This leads to reduced coverage of
$k$-space called undersampling. Special reconstruction techniques are
necessary to recover usable images, since naive Fourier reconstruction
with skipped phase-encoding lines leads to aliasing in the images.
Common parallel imaging techniques include
\glsreset{SENSE}\gls{SENSE}\cite{SENSE},
\glsreset{GRAPPA}\gls{GRAPPA}\cite{GRAPPA}, and
the \glsreset{NLINV}\gls{NLINV} used in this thesis
and described in \cref{sec:nlinv}.
An overview over parallel imaging techniques can be found in
\textcite{parallelMRI}.

\subsection{Radial FLASH}
\label{sec:radial_FLASH}
\subsubsection{FLASH}
\begin{figure}[htbp]
\centering
\begin{subfigure}{0.95\textwidth}
  \centering
 \includegraphics[width=0.9\textwidth]
 {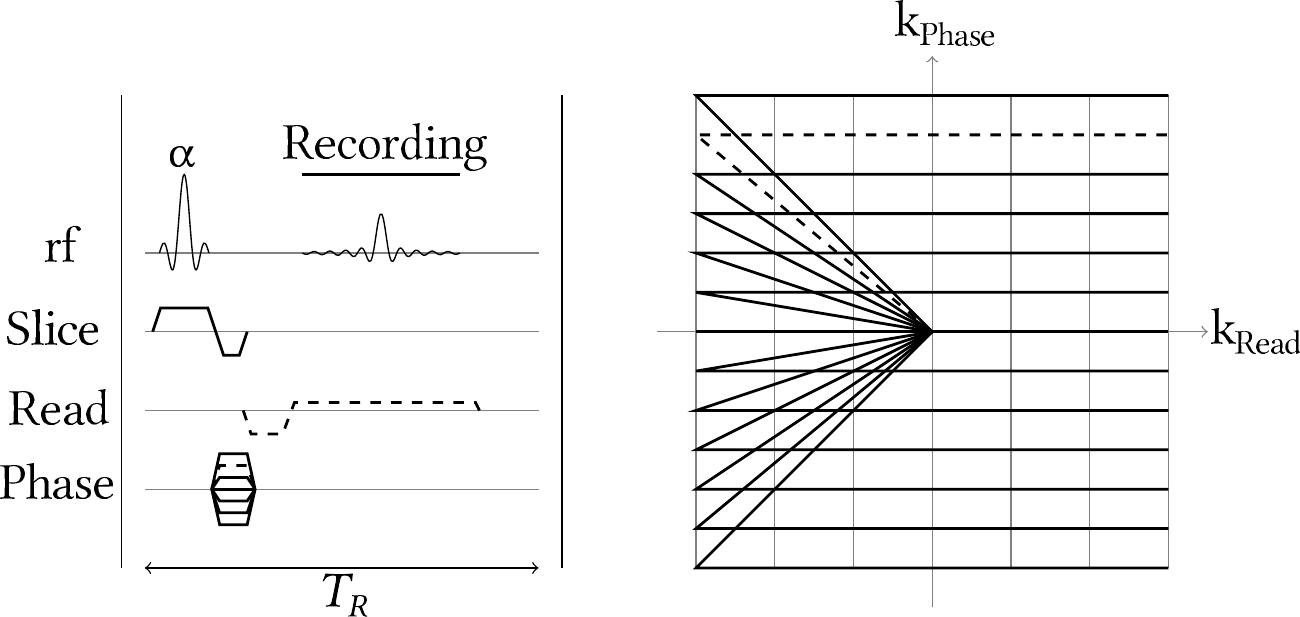}
 \caption{}
 \label{fig:Cartesian}
\end{subfigure}
\vspace{1em}\\
\begin{subfigure}{0.95\textwidth}
  \centering
 \includegraphics[width=0.9\textwidth]
 {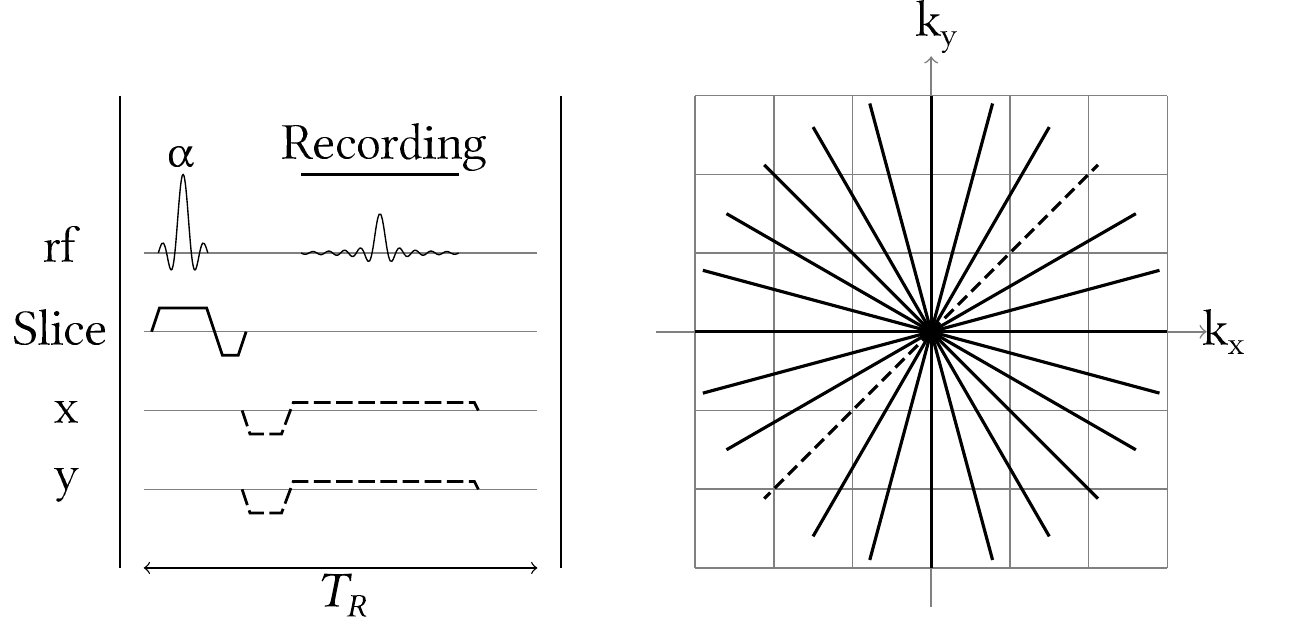}
 \caption{}
 \label{fig:radial}
\end{subfigure}
\caption[]{
Sequence and $k$-space diagrams for \subref{fig:Cartesian} Cartesian
and \subref{fig:radial} radial encoding. The dashed gradient line
in the sequence diagram
corresponds to the dashed $k$-space line in the $k$-space diagram.
Both contain an $\alpha$-pulse and a slice selection gradient.
In \subref{fig:Cartesian} Cartesian encoding, the same
frequency-encoding gradient
is used during readout in all lines, only the phase-encoding
gradient changes. In \subref{fig:radial} radial encoding, both $x$ and
$y$ gradients use frequency encoding, so both are present during readout.
Diagrams adapted from \cite{mueckerphd}.
}
\label{fig:encodings}
\end{figure}
The most important pulse sequence in this thesis is radial
\glsreset{FLASH}\gls{FLASH}.
\gls{FLASH} is a technique for rapid acquisition of \gls{MRI} images,
introduced in \citeyear{FLASHpatent} by \textcite{FLASHpatent}.
The acquisition
scheme described above uses \rf{90}{s}, which maximizes the usable
transversal magnetization for imaging, but also necessitates waiting
for full \gls{T1} recovery before the next excitation. \gls{FLASH}
uses gradient echoes and small angle pulses with flip
angles smaller than \ang{15}.
Combined with continuous imaging, this leads to the development of
a steady state, with the longitudinal magnetization lost to
\gls{rf} excitation being recovered through \gls{T1} relaxation during the
sequence. With continuous imaging, even the transverse phase
coherence does not relax completely. This needs to be rectified, either by
applying strong gradients, called crusher gradients, which completely
dephase the transverse magnetization; or by \gls{rf}-spoiling, which
applies a phase offset onto each excitation pulse, thereby suppressing
the buildup of a steady state of the
transversal magnetization \cite[p. 500f]{haacke}. For fast imaging,
\gls{rf} spoiling is preferred, since it avoids the time-consuming
crusher gradients. With \gls{rf}-spoiled \gls{FLASH}, repetition times
on the order of \SI{2}{\milli\second} are possible.
The steady state which is built up during a \gls{FLASH} acquisition also
dictates the contrast: Since this steady state is \gls{T1}-dependent,
\gls{FLASH}-acquisitions with short repetition times and short echo times
on the order of milliseconds,
like the sequences used for real-time \gls{MRI}, will
show \gls{T1}-weighting, with shorter \gls{T1} appearing brighter.

\subsubsection{Radial Imaging}

Radial encoding is the other difference to current routine \gls{MRI}.
Instead of recording lines on a Cartesian grid in $k$-space, lines
through the $k$-space center are recorded in a pattern similar to spokes
on a wheel.
While it is not necessary, the radial trajectories in this
thesis use equiangular
spacing. A diagram of a radial trajectory is shown in \cref{fig:radial}.
It is immediately obvious that radial trajectories sample
$k$-space non-uniformly. This is a problem in the outer regions
of $k$-space which contain high resolution information: The lower sampling
in this region means that radial trajectories need to measure more lines
to achieve similar resolution to Cartesian trajectories.
The higher sampling
of $k$-space close to the center, however, is an advantage over
Cartesian imaging, since it leads to reduced motion sensitivity. Another
advantage is the absence of a phase-encoding gradient, enabling shorter
echo times and, together with \gls{FLASH}, faster repetition rates.

Radial imaging also allows uniform readout oversampling. This means
faster sampling of the \gls{MRI} signal,
commonly doubling the sampling rate, thereby enlarging the \gls{FOV}
without affecting the resolution. The aliasing artifacts arising
from discrete and finite sampling of $k$-space are hereby moved
further from the region of interest. After reconstruction, the excess
part of the \gls{FOV} is discarded again. Since it is done during
sampling, no additional imaging time is needed, and while
readout oversampling is limited to the frequency-encoding
direction in Cartesian imaging, no such limitation exists
in radial imaging.

Reconstruction also changes with radial encoding: A simple inverse
2D Fourier transform is no longer possible. Instead, there are two main
ways to reconstruct images: Projection reconstruction and gridding.
Projection reconstruction relies on the Fourier slice theorem, which states
that the 1D Fourier transform of lines through the origin of $k$-space
are the projections orthogonal to the line. With this, each measured line can
be Fourier transformed, mapped to an angle, and then reconstructed with
methods like filtered backprojection or other techniques common in
\gls{CT}. Gridding is an alternative approach, where the radial spokes
are resampled onto a Cartesian grid, which is then 2D Fourier transformed.

The most important advantage of radial trajectories is, however, the
possibility of uniform (azimuthal) undersampling:
In Cartesian \gls{MRI}, only the
phase-encoding direction can be undersampled reasonably; undersampling
the frequency-encoding direction does not lead to an appreciable
decrease in measurement time. But the non-uniform sampling
achieved by skipping phase-encoding lines
limits the degree of undersampling
that can be used. With radial sampling, since each spoke
samples both low and high $k$-space regions, measuring fewer spokes
while retaining equiangular spacing leads to uniform undersampling.
For real-time \gls{MRI}, where very high undersampling factors are used,
this uniform undersampling also helps with temporal resolution:
In Cartesian \gls{MRI}, the central $k$-space which contains overall
object shape is only measured once per frame. In radial acquisition, each
spoke measures the $k$-space center,
so each spoke contains equally important data.
Undersampling and related image artifacts is described in more detail
in \cref{sec:artifacts}.
The application of
radial \gls{FLASH} for real-time \gls{MRI} is described, for example,
in \textcite{rt1} and \textcite{rt2}.

\section{Non-linear Inverse Reconstruction}
\label{sec:nlinv}

The most important ingredient for real-time \gls{MRI} is the
\glsreset{NLINV}\gls{NLINV} introduced by \textcite{nlinv} in
\citeyear{nlinv}. 

A closer look at the signal in parallel \gls{MRI} is necessary here.
For a coil array with $N$ coils, the signal in the $j$\textsuperscript{th}
coil is
given by \autocite[p. 25]{mueckerphd}:
\begin{align*}
 s_j(t) = \int \rho(\vect{x}) c_j(\vect{x})\me^{-2\uppi \mi\vect{k}(t)\vect{x}}
  \dd \vect{x} + n(t)
\end{align*}
with the proton density $\rho$, the complex-valued spatial sensitivity
profiles $c_j$, receiver noise $n$, and the used
time-dependent $k$-space trajectory $\vect{k}(t)$.
For \gls{MRI} reconstruction, this has to be discretized and
becomes (\autocite[p. 32]{mueckerphd})
\begin{align}
 \vect{s} = P_k\mathcal{F}C \vect{\rho} + \vect{n}
 \label{eq:discretized_operator}
\end{align}
where $C$ is the multiplication with the
spatial sensitivity profiles (coil profiles),
$\mathcal{F}$ is the discrete Fourier transform, and
$P_k$ is the projection onto the trajectory.

If the spatial sensitivity
profiles (or coil profiles) are known, this can be regarded as a
linear inverse problem (\autocite[Ch. 3.4.2]{mueckerphd})
\begin{align*}
 \vect{y} = A\vect{x} + \vect{n}
\end{align*}
with $A = P_k\mathcal{F}C$. The parallel imaging techniques mentioned
in \cref{sec:MR_principles} can be understood as two-step
approaches, first estimating the
coil profiles from calibration data and then solving this linear inverse
problem.

A problem with the two-step approach is that is does not use the
available data optimally.
The idea of \gls{NLINV} is to jointly estimate image content and
coil profiles, enabling better use of the available data and leading
to higher image quality. Here, \gls{MRI} is understood as a non-linear
operator equation
\begin{align}
 F(x) = y \,,\quad \text{with}\;  x = \begin{pmatrix}
                                   \rho \\
                                   c_1 \\
                                   \vdots \\
                                   c_N
                                  \end{pmatrix}
\label{eq:nonlinear_problem}
\end{align}
with an operator $F$ that maps the unknowns, the
proton density $\rho$ and the coil profile $c_i$ for each of $N$ coils,
to the measured $k$-space data $y$.

For large undersampling factors both the linear and the non-linear
parallel imaging problems become
ill-conditioned, leading to noise amplification. Therefore,
regularization is introduced to curtail the noise amplification.

In this thesis, the \glsreset{IRGNM}\gls{IRGNM} is used to the solve
the non-linear problem. The operator $F$ takes the following form
(compare \cref{eq:discretized_operator}):
\begin{align}
 F\colon x \mapsto \begin{pmatrix}
                   P_k\mathcal{F}{c_1\rho} \\
                   \vdots \\
                   P_k\mathcal{F}{c_N\rho}
                  \end{pmatrix}
\label{eq:operator}
\end{align}
An obvious problem is the insufficient separation between $c_i$ and $\rho$:
multiplying each $c_i$ by any complex function and dividing $\rho$ by the
same function leaves the product unchanged. Even the extreme case of
all proton density information in the coil profiles
$c_i^\prime = c_i^{\vphantom{\prime}}\rho$, $\rho^\prime \equiv 1$
is possible
without further constraints. Since coil profiles are generally smooth,
this problem can be solved with a regularization term penalizing
high $k$-space frequencies in the coil profiles.

As further regularization, the distance to an initial guess can be used.
In the
context of real-time \gls{MRI}, a slightly different approach is useful:
As long as the frame rate is sufficiently high to resolve the dynamics
of the measured object, subsequent frames will be very similar. So instead
of a penalty on the difference to an initial guess, the difference to the
preceding frame is penalized, constituting a form of temporal regularization.

Constraining the reconstruction through prior knowledge in this way is
necessary to recover the data missing because of undersampling.
A more detailed description of the \gls{IRGNM} and its application to
\gls{MRI}
can be found in Chapter~5 of \textcite{mueckerphd} and in \textcite{nlinv}.
The implementation of \gls{NLINV} used in this thesis is described in
\cref{sec:nlinv++}.

\section{Undersampling and Image Artifacts}
\label{sec:artifacts}

Undersampled \gls{MRI} acquisitions are the basis of the real-time \gls{MRI}
used in this thesis, because of the possibility for tremendous speedup.
To understand undersampling, sufficient sampling must be introduced first.
For radial sampling this leads to the following statement:
For a quadratic field of view of length $L$ and a desired resolution
$\Delta x$
determined by the highest sampled frequency
$k_\text{max} = \frac{1}{2\Delta x}$ in $k$-space, the
Nyquist criterion states that the angular spacing $\Delta\phi$
between the spokes has to fulfill \citep[p. 906]{handbook}
\begin{align*}
 k_\text{max}\Delta\phi \leq \frac{1}{L}.
\end{align*}
Or equivalently the number of spokes $n_\text{spokes}$ has to fulfill
\begin{align}
 n_\text{spokes} \geq \uppi Lk_\text{max}
 \label{eq:nyquist}
\end{align}
Compared to the needed number of phase-encoding steps $n_\text{phase}$
in Cartesian sampling
\begin{align}
 n_\text{spokes} = \frac{\uppi}{2}n_\text{phase}
 \label{eq:cart_sampling}
\end{align}
holds,
so approximately \SI{57}{\percent} more excitations are
necessary \citep[p. 906]{handbook}.

\begin{figure}[tbp]
\centering
\begin{subfigure}{0.45\textwidth}
  \centering
 \includegraphics[width=1.0\textwidth]
 {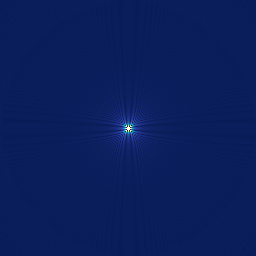}
 \caption{403 spokes (fully sampled)}
 \label{fig:PSFfull}
\end{subfigure}
\hspace{1em}
\begin{subfigure}{0.45\textwidth}
  \centering
 \includegraphics[width=1.0\textwidth]
 {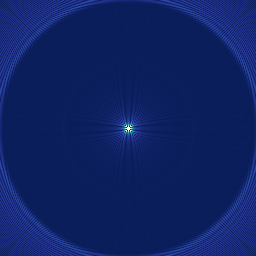}
 \caption{201 spokes}
 \label{fig:PSFhalf}
\end{subfigure}
\\\vspace{1em}
\begin{subfigure}{0.45\textwidth}
  \centering
 \includegraphics[width=1.0\textwidth]
 {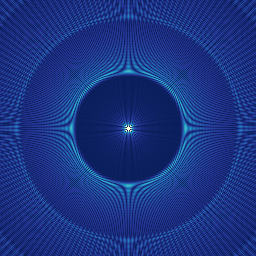}
 \caption{85 spokes}
 \label{fig:PSF85}
\end{subfigure}
\hspace{1em}
\begin{subfigure}{0.45\textwidth}
  \centering
 \includegraphics[width=1.0\textwidth]
 {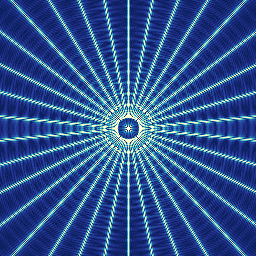}
 \caption{17 spokes}
 \label{fig:PSF17}
\end{subfigure}
\caption[]{Absolute value of simulated point spread functions
for \subref{fig:PSFfull} 403,
\subref{fig:PSFhalf} 201, \subref{fig:PSF85} 85, and \subref{fig:PSF17} 17
spokes. The \gls{FOV} size is $256\times256$. According to
\cref{eq:cart_sampling}, full Nyquist sampling requires more than
402 spokes. 17 and 85 spokes were chosen since those are common values
for the number of spokes per single frame and per full frame in real-time
\gls{MRI} (for a definition of the terminology see \cref{sec:data}).}
\label{fig:PSFs}
\end{figure}

For real-time \gls{MRI}, full Nyquist sampling per frame is too slow,
so far fewer spokes are generally recorded. The effects that this
undersampling has on images can be understood using the
\glsreset{PSF}\gls{PSF}. It is the image of a single point
produced by the imaging process under consideration. So the image of an
extended object is the convolution of this object with the \gls{PSF}.
In \gls{MRI}, the \gls{PSF} is determined by the $k$-space
sampling pattern.
\Cref{fig:PSFfull} shows the \gls{PSF} for a fully sampled radial
acquisition. The \gls{PSF} consists of a central lobe
determining the shape of an imaged point, surrounded
by a region where it is close to zero (called the artifact-free region),
in turn surrounded by artifact-generating side-lobes.
In \cref{fig:PSFfull}, the artifact-free region covers the entire
\gls{FOV}.
Using only half the Nyquist-dictated number of spokes as in
\cref{fig:PSFhalf} leads to larger side lobes covering the
edge of the \gls{FOV}, while even lower numbers of spokes almost
completely eliminate the artifact-free region (\cref{fig:PSF85,fig:PSF17}).

The artifacts generated by this imperfect \gls{PSF} are mostly
streak artifacts, which are generated by each pixel of the object at a
distance determined by the extent of the artifact-free region.
For real-time \gls{MRI}, it is common to record less than $20$ spokes
per frame, leading to a large problem with streak artifacts
(\cref{fig:PSF17}).
Since the image is the convolution of the \gls{PSF} with the object,
high intensity regions of the object will generally lead to more
intense streak artifacts.

The problem of artifacts is partly mitigated by \gls{NLINV}, through
temporal regularization: the \gls{PSF} and
therefore the streak artifacts are different in
subsequent frames, so temporal regularization leads to a dampening.
But since streak artifacts are still a significant problem in some real-time
\gls{MRI} acquisitions, another possible mitigation strategy based
on coil selection is the subject of \cref{sec:Selection}.

\FloatBarrier
\cleardoublepage
\chapter{Methods}
\label{sec:Methods}

\section{Data Acquisition and Processing}
\label{sec:data}
All data used in this thesis was acquired using a
MAGNETOM Primsa\textsuperscript{fit} system
(Siemens Healthcare, Erlangen, Germany) at a field
strength of \SI{3}{\tesla}. For cardiac and abdominal measurements a
32-channel coil array was used, consisting of anterior and posterior
16-channel
arrays. For head measurements a 64-channel head coil was used.
Phantoms used either coil array.

\begin{figure}[htbp]
\centering
 \includegraphics[width=0.8\textwidth]
 {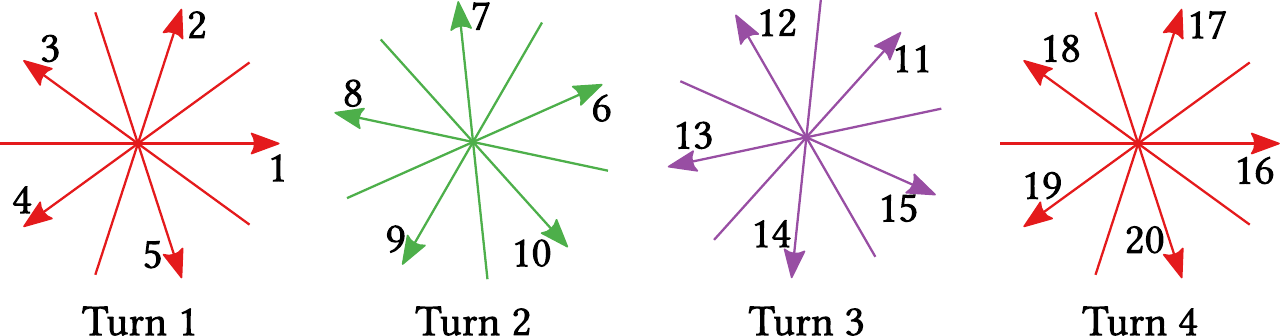}
\caption{
Diagram showing an acquisition with 5 spokes and 3 turns. The numbers next
to the spokes show the order in which they are acquired. Between spokes,
there is an angle of $\sfrac{\ang{360}}{n_\text{spokes}}$
(here: $\sfrac{\ang{360}}{5} = \ang{72}$), between turns
an angle of $\sfrac{\ang{360}}{n_\text{spokes}\cdot n_\text{turns}}$
(here: $\sfrac{\ang{360}}{5\cdot 3} = \ang{24}$).
The $(n_\text{turns}{+}1)$\textsuperscript{th}
turn is again identical to the first turn.
}
 \label{fig:turns}
\end{figure}

A radial \gls{FLASH} sequence developed in the insitute was used for
acquisition. This sequence uses a turn-based pattern, where a set of
$n_\text{spokes}$ spokes is acquired for each frame. For the subsequent
frame, this pattern is rotated by
$\sfrac{2\pi}{n_\text{turns}\cdot n_\text{spokes}}$, so that after
$n_\text{turns}$ frames the patterns overlap again. This is illustrated
in \cref{fig:turns}. $n_\text{turns}$ frames taken together, containing
all spoke orientations, are called a \emph{full frame}.

Acquisition parameters for the datasets used in this thesis can be found in
\cref{tab:sequences} in \cref{sec:sequences}, while
descriptions of the datasets are gathered in \cref{tab:descriptions}.
Dataset labels are typeset in
small capitals (e.g. \data{Head1}) so that they can be
quickly identified.

\raggedbottom

\subsection{Pre- and Postprocessing}
\label{sec:pre_postprocessing}
There are several pre- and postprocessing steps applied to the measured data.
First is a gradient delay correction using the first full frame:
due to imperfect gradient timing in
the \gls{MRI} system, the actually measured $k$-space trajectory
deviates from the expected trajectory. This can partly be corrected by
gradient delay correction. The procedure is described in detail in
Chapter 4 of \textcite{muntenbergerphd}.
Then, the data is compressed by a \glsreset{PCA}\gls{PCA} to fewer channels.
This procedure is described in \cref{sec:Compression}.
The corrected and compressed data is then interpolated on a
square Cartesian grid
in a procedure called \emph{gridding}, described, for example, in Chapter
13.2 of \textcite{handbook}. For better accuracy of the interpolation,
the Cartesian grid is chosen with a $1.5$ times higher sample density than the
originally sampled $k$-space data. This is called \emph{overgridding}.
The gridded frames are then reconstructed using \gls{NLINV}.

The postprocessing steps depend on the application. For anatomical
real-time \gls{MRI}, a pixelwise temporal median filter is applied: Each
pixel in frame $n$ is replaced by the median of the same pixel in the frames
$n \pm \floor{\frac{n_\text{turns}}{2}}$. A non-local means filter for
image denoising, described in \textcite{jakobfilter}, is applied next.

For other applications, different postprocessing steps are needed.
For phase-contrast flow imaging the phase difference map needs to be calculated,
for example.

All pre- and postprocessing steps are available in two implementations:
a \matlab\footnote{\matlab is a registered trademark of The MathWorks,
Massachusetts, USA.}
implementation and integrated in \nlinvpp.
The \matlab implementation is used for testing and implementing
new techniques, while \nlinvpp can be used directly on the
\gls{MRI} scanner.

\subsection{nlinv++}
\label{sec:nlinv++}
\iffalse
\nlinvpp is a \texttt{C++} program written in the institute,
described in \textcite{nlinv++}.
\else
\nlinvpp is a \texttt{C++} program written in the institute.
\fi
It can be used online
(i.e. on the \gls{MRI} scanner) and offline. The online version is running
as a ``bypass server'', so instead of forwarding the data to the
vendor-supplied reconstruction pipeline, it is send to the custom
reconstruction pipeline instead. The reconstructed images are then written
back to the scanner, yielding seamless integration.
Offline, \nlinvpp can be used to reconstruct previously measured data again.

\nlinvpp is itself implemented as a pipeline. It consists of five
pipeline stages, each running in its own thread with an additional thread
for the controlling stage. The five pipeline stages are: \texttt{data\-source},
\texttt{pre\-pro\-ces\-sor}, \texttt{re\-con\-struc\-tor},
\texttt{post\-pro\-ces\-sor}, and
\texttt{data\-sink}. The
stages are connected by channels, which are implemented as thread-safe
type-erased lists, so any kind of message may be
passed along the pipeline. Each pipeline stage decides on the appropriate
action based on the type of this
message\interfootnotelinepenalty=10000\footnote{The type is only erased inside the
channels, it is
recovered in the pipeline stages to decide on the action to be taken.}.
Since the \gls{NLINV} algorithm contains Fourier transforms and
matrix-products of large matrices, the \texttt{reconstructor} stage uses
\glspl{GPU} to accelerate computation.

\nlinvpp is running on dedicated Supermicro SuperServer 4027GR-TR
system with the Ubuntu 14.04 operating system,
2x Intel Xeon Ivy Bridge E5-2650 main processors,
8x NVIDIA GTX Titan Black (Kepler GK110) \glspl{GPU} as accelerators,
and \SI{128}{\gibi\byte} main memory.

The setup is as follows: the controlling \texttt{main}
thread starts up and sets up the configuration. It then starts up each
of the pipeline stages:
\paragraph{\texttt{datasource}}
It reads in the raw measured data, either from a file or from a pipe connected
to the scanner. It buffers data until enough data for one frame has been
read and then sends it on.

\paragraph{\texttt{preprocessor}}
It receives the raw data from the \texttt{datasource} one frame at a time.
It uses the first full frame as calibration data to calculate the gradient
delay values and the
coil compression matrix. It applies gradient delay correction, coil compression
and gridding to each frame. Coil compression is discussed in more detail in
\cref{sec:Compression}.

\paragraph{\texttt{reconstructor}}
The \texttt{reconstructor} is the only stage that is itself multithreaded,
commonly using up to 4 threads. Each thread is identical, and each
reads its input from the common channel coming from the
\texttt{preprocessor}. Since 8 \glspl{GPU} are available, each
of $n_\text{threads}$ \texttt{reconstructor} threads
distributes its data onto $\floor{\frac{8}{n_\text{threads}}}$
\glspl{GPU}. Each thread independently reconstructs its frame
and places it in the channel to the \texttt{postprocessor}.

\paragraph{\texttt{postprocessor}}
Receives reconstructed frames from the \texttt{reconstructor} and applies
appropriate postprocessing to them. In most cases,
that means temporal median and non-local means filtering. For phase-contrast
flow \gls{MRI} for example,
it includes calculation of the phase-contrast map.

\paragraph{\texttt{datasink}}
Receives postprocessed images and writes them to file. Since the
\texttt{re\-con\-struc\-tor} is multithreaded, it might receive frames
out-of-order.
In that case, it buffers the out of order frames while waiting for the next
in-order frame.

\hspace{0em}\\
Once all data has been read, the \texttt{datasource} produces a
\texttt{finished} message. Each subsequent pipeline stage that
receives this message produces its own \texttt{finished} message
and exits as well. The \texttt{datasink}, after having written
all output data, produces a \texttt{finished} message for the
controlling \texttt{main} thread. This final message acts as a
control that the reconstruction finished successfully.

\section{Data Visualization}

The 2D images in this thesis are created using
\texttt{arrayShow}\footnote{\url{https://github.com/BiomedNMR/arrShow},
\texttt{arrayShow} is a \matlab image viewer for multidimensional arrays,
with a focus on \gls{MRI} images.}.
Unless otherwise indicated, images are individually windowed.
This ensures that comparisons between images are not unduly influenced by,
for example, differences in total signal content.
Furthermore, since the data analyzed in this thesis stems from
real-time \gls{MRI} acquisitions, they are properly understood as
movies rather than still images. Therefore, representative individual frames
were selected for this thesis.

\begin{figure}[tbp]
\centering
\begin{subfigure}{0.45\textwidth}
  \centering
 \includegraphics[width=1.0\textwidth]
 {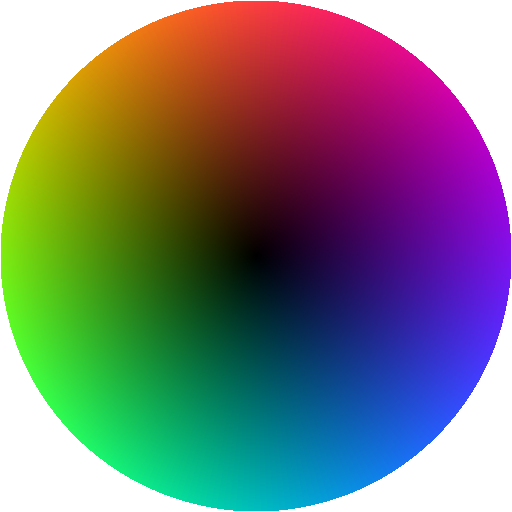}
 \caption{Complex colormap}
 \label{fig:colormap_complex}
\end{subfigure}
\hspace{1em}
\begin{subfigure}{0.45\textwidth}
  \centering
 \includegraphics[width=1.0\textwidth]
 {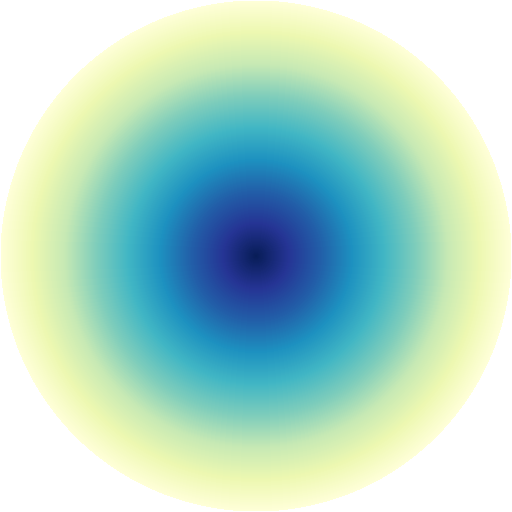}
 \caption{Magnitude colormap}
 \label{fig:colormap_magnitude}
\end{subfigure}
\caption[]{Colormaps used in this
thesis. Both \subref{fig:colormap_complex} and \subref{fig:colormap_magnitude}
show the complex unit disk. \subref{fig:colormap_complex} shows the colormap
used for directly representing complex-valued data,
with hue indicating phase angle and
brightness indicating absolute value. \subref{fig:colormap_magnitude} shows
the colormap used for representing the magntidue of complex data.
The colormap in \subref{fig:colormap_complex} is the standard colormap
of \texttt{arrayShow} for complex data, while the colormap in
\subref{fig:colormap_magnitude} is the \texttt{YlGnBu\_r}
(``yellow, green, blue, reversed'') colormap from the
\texttt{matplotlib} project, optionally available in \texttt{arrayShow}.
Purely real data is visualized with a grayscale colormap.}
\label{fig:colormaps}
\end{figure}

\Cref{fig:colormaps} shows the colormaps used in this thesis.
For ease of identification, the colormap in \cref{fig:colormap_complex}
is used to directly represent complex data, while the
colormap in \cref{fig:colormap_magnitude} is used when
showing the magnitude of complex data. A plain grayscale colormap is
used for purely real data.

The comparison of different methods in this thesis is done by visual
inspection of images, or by inspecting their difference. This is done
since no generally accepted criterion for quantification of image quality
exists in \gls{MRI}, and since medical images are traditionally
interpreted through visual inspection by physicians.
Furthermore, the postprocessing filters
described in \cref{sec:pre_postprocessing} are not used on the
images, so that the comparison is not unduly influenced by them.

\FloatBarrier
\cleardoublepage
\chapter{Coil Compression}
\label{sec:Compression}
A challenge in modern \gls{MRI} is the large amount of measurement data that
can be acquired with current systems. Arrays of 32 or 64 coils are in routine
use today, with even larger numbers of coils in consideration\cite{128ch}.
The use of very fast imaging techniques in real-time \gls{MRI} exacerbates this
problem. One advantage of multiple receiver coils is the possibility
of parallel imaging for improved image quality and faster acquisition. The
drawback is increased computation time.

A way to address this problem is coil compression, that means finding a
smaller approximation of the data that still captures as much of the contained
information as possible. One possibility is combining the measured coils into
a smaller set of ``virtual coils'', which is the approach taken here. One
very fast approach is the use of linear combinations of coils, calculated from
some
initial calibration data. This enables the use of the once calculated
compression
operator on new data as it is coming in, in a way suitable for online use.
Furthermore, the channel compression can be represented as a matrix product
with
the incoming data, which is advantageous because of the wide availability of
highly optimized matrix-matrix-product routines.

Especially the ease of online use is the reason why a linear method was chosen
for the current \nlinvpp implementation, namely \gls{PCA}.

\section{Principal Component Analysis}
\label{sec:compression_PCA}
Principal component analysis (see \cite{jolliffe2006PCA} for a general
introduction) is a technique for dimensionality
reduction and feature extraction used in statistics.
Here, the
dimensionality reduction is the important characteristic.

\subsection{The PCA Algorithm}
\label{sec:PCA_algo}
The input of the \gls{PCA} algorithm is the ungridded raw data of the
first full frame from scanner.
This data consists of the measured samples for each line for each coil,
so it is an array of dimensions
$ n_\text{coils} \times n_\text{S} \times n_\text{samples}$, which for
\gls{PCA} is treated as a matrix of dimensions
$n_\text{coils} \times n_\text{S} \cdot n_\text{samples}$.
Here, $n_\text{S} = n_\text{spokes} \cdot n_\text{turns}$ is the number
of spokes in a full frame.

\begin{enumerate}
 \item Take the input matrix $A$ of size
    $n_\text{coils} \times n_\text{S} \cdot n_\text{samples}$
    and calculate the $n_\text{coils} \times n_\text{coils}$ matrix
    $\text{cov}(A) \defd A^*\cdot A$, where $^*$ is the conjugate transpose
    operation. $\text{cov}(A)$ is related to the covariance matrix of A, but
    differs in two important aspects. First, the covariance matrix requires
    the columns of $A$ to have zero mean. This is not guaranteed for
    \gls{MRI} data. The column means are,
    however, close enough to zero to have negligible influence on the method.
    Second, the covariance matrix needs to be scaled by
    $\sfrac{1}{n_\text{samples}}$, which $\text{cov}(A)$ is not.\footnote{So
    $\text{cov}(A)$ can only be proportional to the covariance matrix. However,
    this is sufficient, as will be explained later.}

  \item Find an eigenvalue decomposition of $\text{cov}(A)$, that means
    a unitary transformation matrix $I$ and a diagonal matrix $\Lambda$
    so that
    $\text{cov}(A) = I^* \cdot \Lambda \cdot I$. $\Lambda$ contains
    the eigenvalues of $\text{cov}(A)$ on its diagonal, and all eigenvalues
    are real since $\text{cov}(A)$ is a normal matrix\footnote{A matrix
    $M$ is normal if $M\cdot M^* = M^*\cdot M$, which is obviously
    fullfilled for $\text{cov}(A)$}.

  \item Jointly permute $I$ and $\Lambda$ so that the eigenvalues on the
    diagonal of $\Lambda$ are in descending order.

  \item Use the first $n_\text{pc}$ columns of $I$ as the compression matrix,
  where
    $n_\text{pc}$ is the desired number of principal components.
\end{enumerate}
This method can be understood in the following way: By applying $I$ to the
input data it is transformed into a different coordinate system. In this
system, the diagonal entries of $\Lambda$ are a measure of the variation
contained in each direction. The variation contained in the
$i$\textsuperscript{th} direction
can be calculated as $\sfrac{\Lambda_{ii}}{\text{tr}(\Lambda)}$, where
$\text{tr}()$ is the trace. So the first $n_\text{pc}$ columns of $I$ are a
transformation matrix which retains
$\sfrac{\sum_{i=0}^{n_\text{pc}}\Lambda_{ii}}{\text{tr}(\Lambda)}$ of the
variation
in $A$. This normalization with $\text{tr}(\Lambda)$ is the reason why it is
enough for the eigenvalues of $\Lambda$ to be proportional to the covariances.

\subsection{Application in real-time MRI}
For real-time \gls{MRI}, the \gls{PCA} algorithm is applied as a
preprocessing step. In the
\texttt{pre\-pro\-ces\-sor} stage (see \cref{sec:nlinv++} for a
discussion of
the pipeline stages), the data of the first full frame is used as input for
the algorithm. In the current implementation, $n_\text{pc} = 10$, that is
the first
10 principal components are used. This is a heuristically obtained,
conservative value \cite[Ch. 6.3]{szhangphd}.
Images for comparison between $n_\text{pc} = 10$ and
$n_\text{pc} = n_{\text{coils}}$ are shown in \cref{fig:npc_comparison}.

\begin{figure}[tbp]
\centering
\foreach \index in {full,10} {%
\begin{subfigure}{0.38\textwidth}
  \centering
 \includegraphics[width=0.95\textwidth]
 {figures/matlab_figs/T3506_phantom/T3506_PCA_\index.png}
 \caption{$n_\text{pc} = $ \index}
 \label{fig:npc_comparison_T3506_\index}
\end{subfigure}\hspace{0.1\textwidth}
}\hspace{-0.1\textwidth}\vspace{1em}\\

\foreach \index in {full,10} {%
\begin{subfigure}{0.38\textwidth}
  \centering
 \includegraphics[width=0.95\textwidth]
 {figures/matlab_figs/T6954_hcmh/T6954_PCA_\index.png}
 \caption{$n_\text{pc} = $ \index}
 \label{fig:npc_comparison_T6954_\index}
\end{subfigure}\hspace{0.1\textwidth}
}\hspace{-0.1\textwidth}\vspace{1em}\\

\foreach \index in {full,10} {%
\begin{subfigure}{0.38\textwidth}
  \centering
 \includegraphics[width=0.95\textwidth]
 {figures/matlab_figs/T18907_cardiac_hcmh/T18907_PCA_PD_\index.png}
 \caption{$n_\text{pc} = $ \index}
 \label{fig:npc_comparison_T18907_\index}
\end{subfigure}\hspace{0.1\textwidth}
}\hspace{-0.1\textwidth}

\caption[]{Comparison of using $10$ principal components
vs. using the full dataset for a water phantom (\data{Phan1})
\subref{fig:npc_comparison_T3506_full}--\subref{fig:npc_comparison_T3506_10},
a sagittal slice through a human head (\data{Head1})
\subref{fig:npc_comparison_T6954_full}--\subref{fig:npc_comparison_T6954_10},
and a two-chamber view of a human heart (\data{Heart1})
\subref{fig:npc_comparison_T18907_full}--\subref{fig:npc_comparison_T18907_10}.
The difference in image quality is negligible. As an example
\cref{fig:npc_diff}
shows the magnitude of the complex difference beween
\subref{fig:npc_comparison_T6954_full} and
\subref{fig:npc_comparison_T6954_10}.
}
\label{fig:npc_comparison}
\end{figure}

\begin{figure}[tbp]
\centering
  \centering
 \includegraphics[width=0.42\textwidth]
{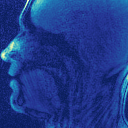}
 \caption{Magnitude of the complex difference between
 \cref{fig:npc_comparison_T6954_full,fig:npc_comparison_T6954_10}, shown
 as an example. Since the
 image is mostly noise with only little structure in the high intensity
 regions, it can be concluded that the \gls{PCA} compression to 10 coils
 does not impact image quality too much.
 }
 \label{fig:npc_diff}
\end{figure}

There are several shortcoming of this approach:
\begin{enumerate}
 \item It is independent of the image under consideration. Some images,
    especially of phantoms, can naturally be described by fewer components.
    Furthermore, the number of imaging coils is not taken into consideration.
    Using 10 principal components will have vastly different results when
    using e.g. a 12-channel compared to a 64-channel head coil.

  \item 10 only has 2 and 5 as factors.
    When distributing coil data onto
    accelerators for reconstruction (as described in \cref{sec:nlinv++}),
    this might lead to some accelerators containing more coil data to calculate
    than others. This will slow down the accelerators having to do more work,
    and through synchronization between the accelerators slow down the whole
    reconstruction.

  \item \gls{PCA} finds ``virtual coils'' (directions) of highest variance.
    But this
    is not necessarily optimal for \gls{MRI}: \gls{PCA} will favor areas of
    high signal
    intensity at the expense of areas of lower intensity, creating not only
    non-uniform image appearance, but it can also lead to higher noise levels,
    since the coils with high sensitivity in the regions of low
    intensity will be weighted less. This effect, of course, is less
    pronounced at higher numbers of principal components.
\end{enumerate}

Therefore, an optimized coil compression algorithm was investigated in this
thesis, in order to lower the number of necessary principal components and
thereby improve reconstruction speeds.


\FloatBarrier
\section{Optimized Combination}
\label{sec:compression_optimized}

In order to address these shortcomings of \gls{PCA} for \gls{MRI},
\citet{optimized_combination} have introduced an optimized
coil compression algorithm for \gls{SENSE} \gls{MRI} \cite{SENSE}. A \matlab
implementation adapted for
real-time \gls{MRI} was already available in the institute\footnote{Written by
Soeren Wolfers as part of an internship program. Because of the limited
duration of the internship, no performance evaluation
was done at that time.}, and was used for evaluation of the method.

\begin{figure}[tbp]
\centering
\begin{subfigure}{0.45\textwidth}
  \centering
 \includegraphics[width=0.95\textwidth]
 {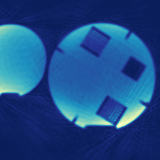}
 \caption{}
 \label{fig:optimized_PD}
\end{subfigure}\hspace{0.1\textwidth}%
\begin{subfigure}{0.45\textwidth}
  \centering
 \includegraphics[width=0.95\textwidth]
 {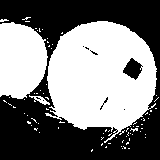}
 \caption{}
 \label{fig:optimized_ithresh}
\end{subfigure}%
\vspace{1em}\\
\begin{subfigure}{0.45\textwidth}
  \centering
 \includegraphics[width=0.95\textwidth]
 {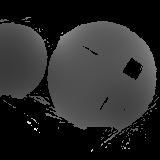}
 \caption{}
 \label{fig:optimized_cadj}
\end{subfigure}\hspace{0.1\textwidth}%
\begin{subfigure}{0.45\textwidth}
  \centering
 \includegraphics[width=0.95\textwidth]
 {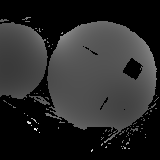}
 \caption{}
 \label{fig:optimized_cthresh}
\end{subfigure}%

\caption[]{
Visualization of different steps of the optimized algorithm.
\subref{fig:optimized_PD}: magnitude of the input proton density $R$.
\subref{fig:optimized_ithresh}--\subref{fig:optimized_cthresh}:
weighting matrix $W$ after \subref{fig:optimized_ithresh} \SI{5}{\percent}
thresholding of the proton density, \subref{fig:optimized_cadj}
normalization with the combined sensitivities,
\subref{fig:optimized_cthresh} \SI{5}{\percent}
thresholding of the combined sensitivities. For this dataset
this last step leaves $W$ unchanged, since the combined sensitivities
pass the threshold at every nonzero pixel.
}
\label{fig:optimized}
\end{figure}

\begin{figure}[tbp]
\centering
  \centering
 \includegraphics[width=0.92\textwidth]
 {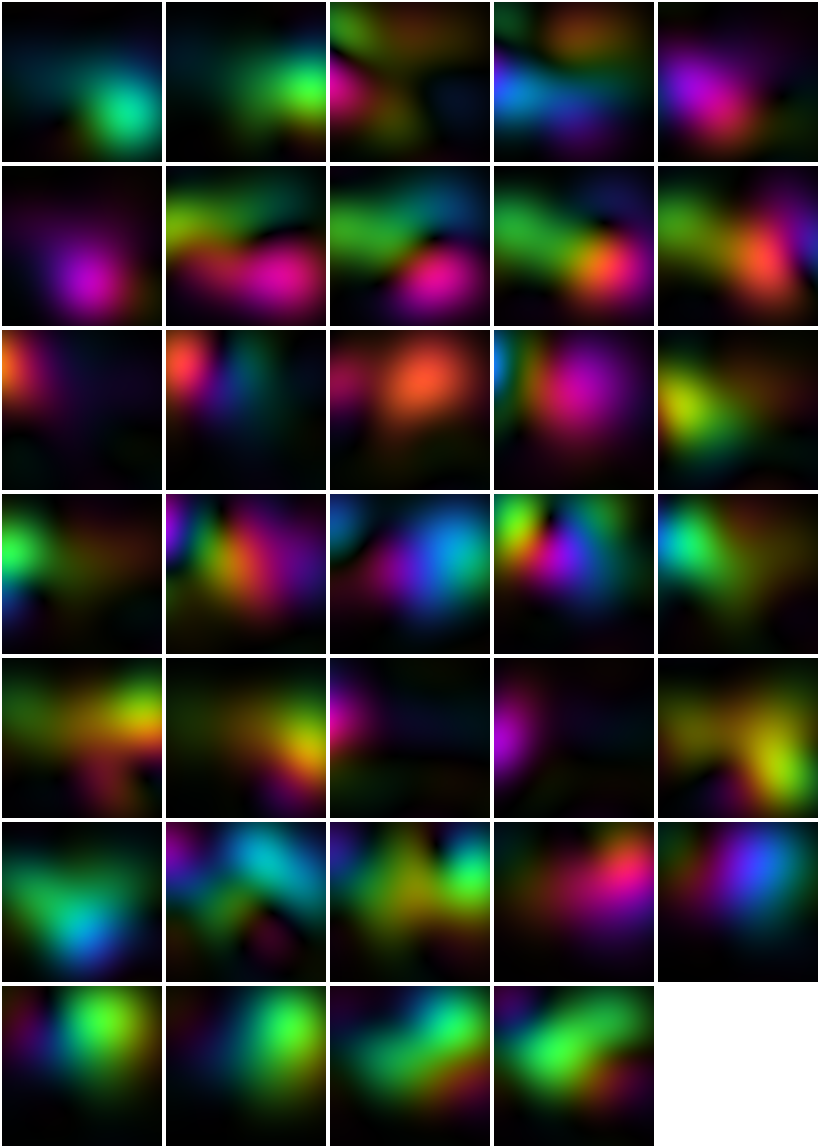}
 \caption{
 Complex-valued input coil profiles $C$ for the
 optimized algorithm.
 }
 \label{fig:optimized_C}
\end{figure}

\begin{figure}[tbp]
\centering
  \centering
 \includegraphics[width=0.92\textwidth]
 {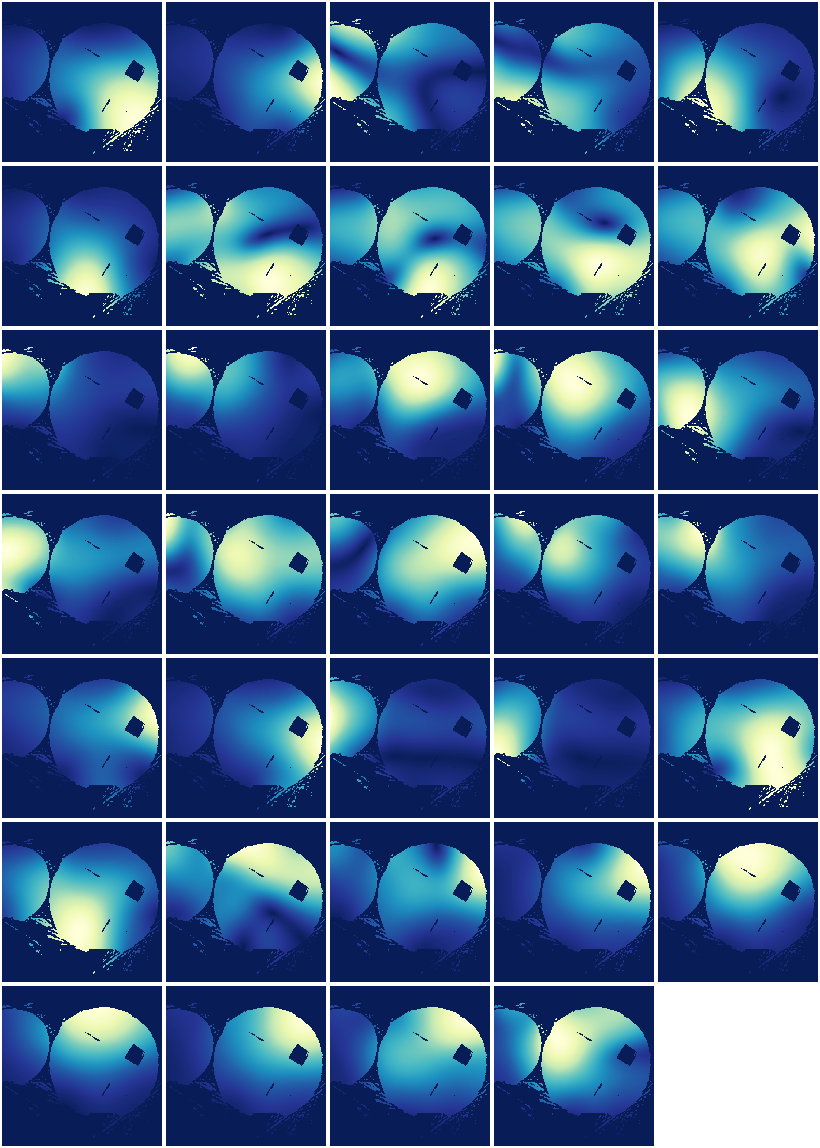}
  \caption{Weighted coil profiles $C^w$ of the optimized algorithm,
  used as input for regular \gls{PCA}.
  }
 \label{fig:optimized_weighted_C}
\end{figure}

The optimized algorithm tries to overcome some of
the shortcomings of plain \gls{PCA},
especially the non-uniform image appearance and the possible noise enhancement
by using the coil profiles as weighting.
\Cref{fig:optimized,fig:optimized_C,fig:optimized_weighted_C} show the
different steps of the algorithm
applied to a water phantom dataset (\data{Phan1}).
\begin{enumerate}
 \item Input to the optimized algorithm are the
    $n_\text{pixels}\times n_\text{pixels}$ proton density $R$
    (\cref{fig:optimized_PD}) and the
    $n_\text{pixels}\times n_\text{pixels}\times n_\text{coils}$
    coil sensitivity profiles $C$ (\cref{fig:optimized_C})
    of a finished \gls{NLINV} run. Generate a
    $n_\text{pixels}\times n_\text{pixels}$
    weighting matrix $W$ with $\forall\, i,j\,W_{ij} = 1$.

  \item Only consider pixels of moderately high image intensity. The threshold
    for this was set at \SI{5}{\percent} of the maximum intensity: Set
    $W$ to zero at
    $\{\,(i,j) \mid R_{ij} < \min(R) + 0.05 \cdot (\max(R)-\min(R))\,\}$
    (\cref{fig:optimized_ithresh}).

  \item Calculate the combined sensitivity matrix $S$:
    $S_{ij} = \sqrt{\sum_{c=1}^{n_\text{coils}}\|C_{ijc}\|^2}$.

  \item Normalize coil sensitivity in each pixel, i.e. divide
    $W$ at each point $(i,j)$ by $S_{ij}$
    (\cref{fig:optimized_cadj}).

  \item Only considers pixels of moderately high combined coil sensitivity.
    Here, the same threshold of \SI{5}{\percent} of the maximum was chosen:
    Set $W$ to zero at
    $\{\,(i,j) \mid S_{ij} < \min(S)) + 0.05 \cdot (\max(S)-\min(S))\,\}$
    (\cref{fig:optimized_cthresh}).

  \item Calculate weighted coil sensitivities:
    $C^w_{ijc} = C_{ijc} \cdot W_{ij}$ (\cref{fig:optimized_weighted_C})

  \item Transpose and interpret the
  $n_\text{pixels}\times n_\text{pixels}\times n_\text{coils}$
  weighted coil sensitivities matrix $C^w$ as a
  $n_\text{coils} \times n_\text{pixels}^2$ matrix and use the PCA
  algorithm described in \cref{sec:PCA_algo}.
\end{enumerate}
This algorithm takes some insights into account. First, by using the proton
density and coil sensitivity profiles, it is possible to exclude irrelevant
pixels in the calculation. Furthermore, to generate more uniform image
intensity and noise levels, the coil profiles are normalized prior to PCA
compression. A disadvantage of this method is the need for the output
of a finished \gls{NLINV} run.

\FloatBarrier
\section{Performance Evaluation}
Since the optimized algorithm has not already been added into \nlinvpp,
the performance evaluation was strictly done offline, using \matlab.
Routines implementing data reading and preprocessing (including gradient
delay correction and gridding) were already available. For the reconstruction,
a modified version of \nlinvpp was written as part of this thesis,
which allows the input of
data preprocessed in \matlab and only uses the \texttt{reconstructor} and
\texttt{datasink} stages described in \cref{sec:nlinv++}.
This was done
to prevent subtleties in the reconstruction from influencing the results
of the evaluation.

\begin{figure}[htbp]
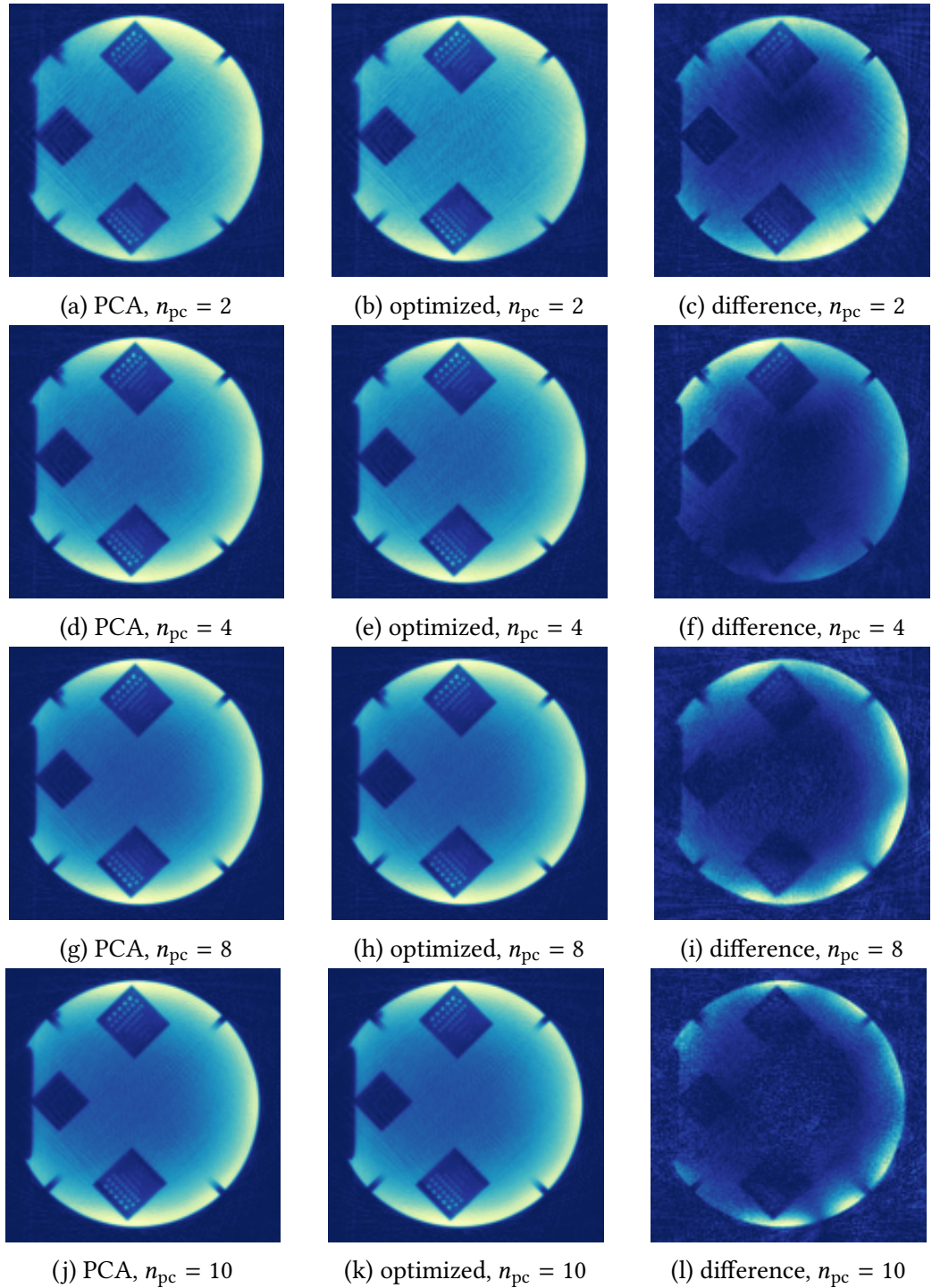

\centering
\foreach \index in {2,4,8,10} {%
\begin{subfigure}{0.3\textwidth}
  \centering
 \includegraphics[width=0.90\textwidth]
 {figures/matlab_figs/T16677_structure_phantom_9spk/T16677_PCA_PD_\index.png}
 \caption{PCA, $n_\text{pc} = \index$}
 \label{fig:compr_T16677_PCA\index}
\end{subfigure}\hspace{0.01\textwidth}
\begin{subfigure}{0.3\textwidth}
  \centering
 \includegraphics[width=0.90\textwidth]
 {figures/matlab_figs/T16677_structure_phantom_9spk/T16677_Soeren_PD_\index.png}
 \caption{optimized, $n_\text{pc} = \index$}
 \label{fig:compr_T16677_Soeren\index}
\end{subfigure}\hspace{0.01\textwidth}
\begin{subfigure}{0.3\textwidth}
  \centering
 \includegraphics[width=0.90\textwidth]
 {figures/matlab_figs/T16677_structure_phantom_9spk/T16677_Diff_PD_\index.png}
 \caption{difference, $n_\text{pc} = \index$}
 \label{fig:compr_T16677_Diff\index}
\end{subfigure}
}
\caption[]{
Dataset \data{Phan2} compressed with \gls{PCA} (left column),
the optimized combination (middle) and their difference (right column).
While the
\gls{SNR} increases with increasing number of principal components, the
visual difference between \gls{PCA} and the optimized combination
is negligible.
}
\label{fig:compr_T16677_phantom}
\end{figure}

\begin{figure}[htbp]
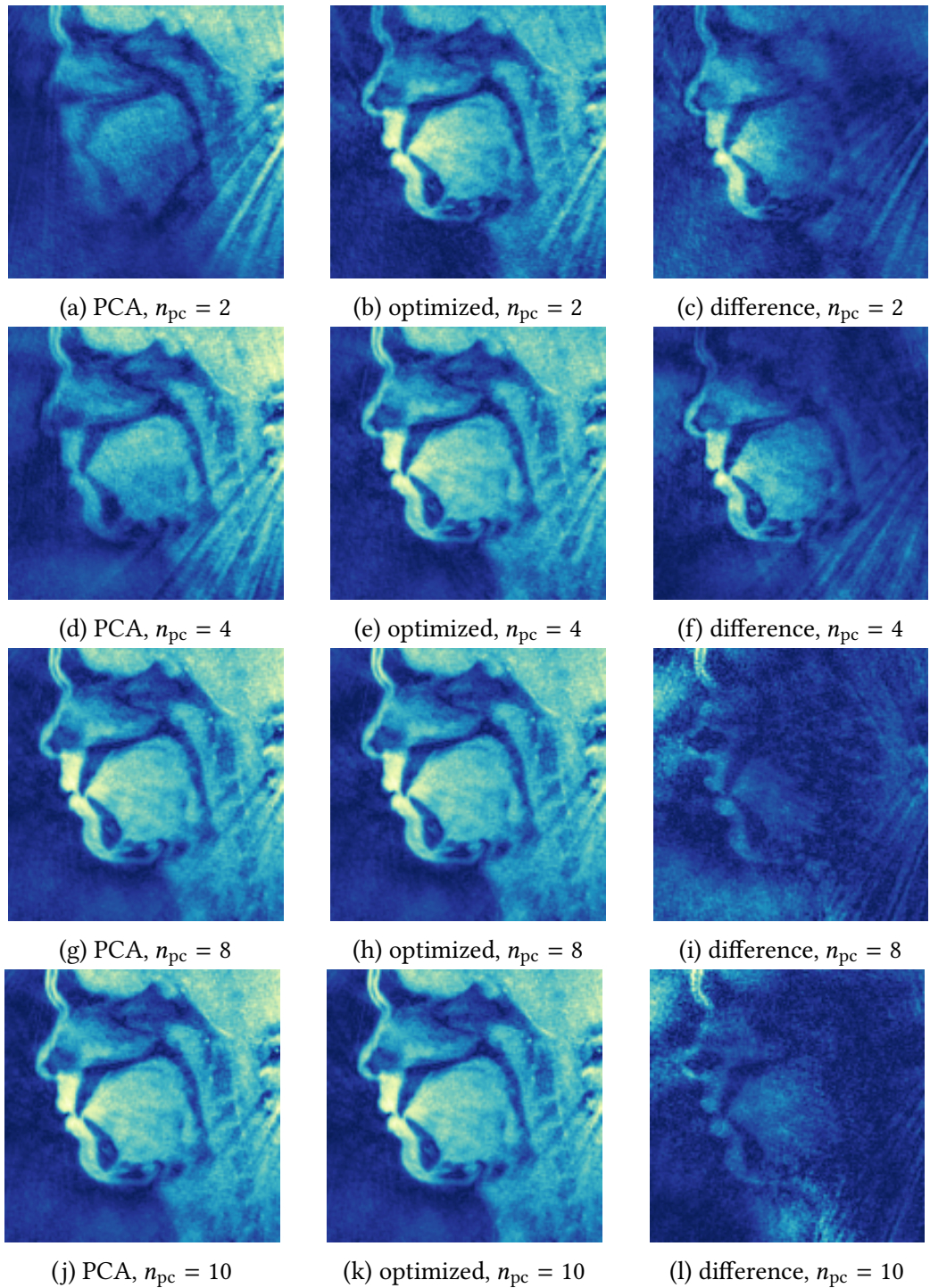

\centering
\foreach \index in {2,4,8,10} {%
\begin{subfigure}{0.3\textwidth}
  \centering
 \includegraphics[width=0.90\textwidth]
 {figures/matlab_figs/T17086_brass_oct/T17086_PCA_PD_\index.png}
 \caption{PCA, $n_\text{pc} = \index$}
 \label{fig:compr_T17086_brass_oct_PCA\index}
\end{subfigure}\hspace{0.01\textwidth}
\begin{subfigure}{0.3\textwidth}
  \centering
 \includegraphics[width=0.90\textwidth]
 {figures/matlab_figs/T17086_brass_oct/T17086_Soeren_PD_\index.png}
 \caption{optimized, $n_\text{pc} = \index$}
 \label{fig:compr_T17086_brass_oct_Soeren\index}
\end{subfigure}\hspace{0.01\textwidth}
\begin{subfigure}{0.3\textwidth}
  \centering
 \includegraphics[width=0.90\textwidth]
 {figures/matlab_figs/T17086_brass_oct/T17086_Diff_PD_\index.png}
 \caption{difference, $n_\text{pc} = \index$}
 \label{fig:compr_T17086_brass_oct_Diff\index}
\end{subfigure}
}
\caption[]{
Dataset \data{Head3} compressed with \gls{PCA} (left column),
the optimized combination (middle) and their difference (right column).
For 2 and
4 principal components, the optimized combination yields
preferable images, even reducing the severity of streak
artifacts in the neck region between
\subref{fig:compr_T17086_brass_oct_PCA4}
and \subref{fig:compr_T17086_brass_oct_Soeren4}.
For 8 and 10 principal components, the difference
is negligible.
}
\label{fig:compr_T17086_brass_oct}
\end{figure}

\begin{figure}[htbp]
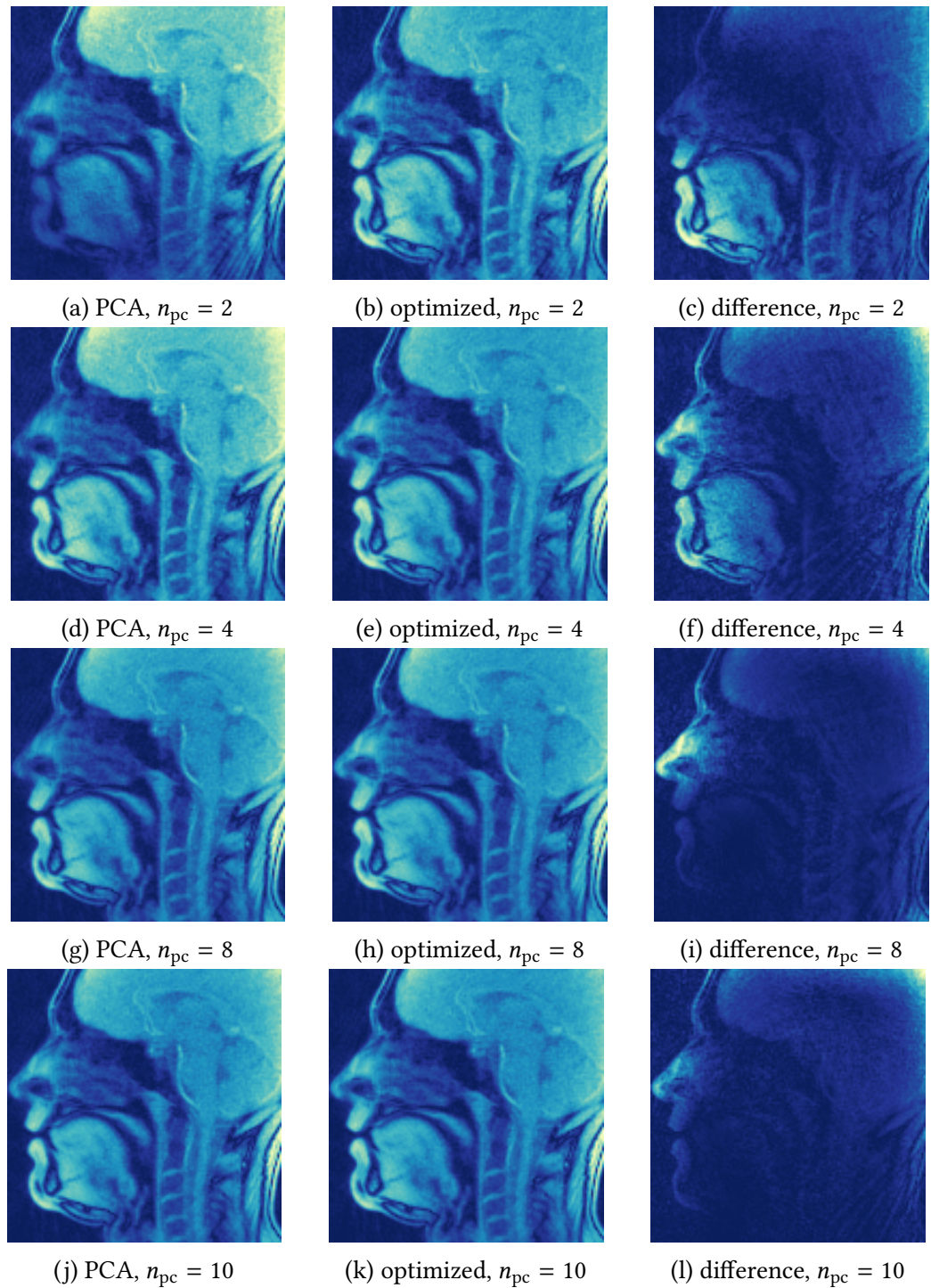

\centering
\foreach \index in {2,4,8,10} {%
\begin{subfigure}{0.3\textwidth}
  \centering
 \includegraphics[width=0.90\textwidth]
 {figures/matlab_figs/T6954_hcmh/T6954_PCA_PD_\index.png}
 \caption{PCA, $n_\text{pc} = \index$}
 \label{fig:compr_T6954_hcmh_PCA\index}
\end{subfigure}\hspace{0.01\textwidth}
\begin{subfigure}{0.3\textwidth}
  \centering
 \includegraphics[width=0.90\textwidth]
 {figures/matlab_figs/T6954_hcmh/T6954_Soeren_PD_\index.png}
 \caption{optimized, $n_\text{pc} = \index$}
 \label{fig:compr_T6954_hcmh_Soeren\index}
\end{subfigure}\hspace{0.01\textwidth}
\begin{subfigure}{0.3\textwidth}
  \centering
 \includegraphics[width=0.90\textwidth]
 {figures/matlab_figs/T6954_hcmh/T6954_Diff_PD_\index.png}
 \caption{difference, $n_\text{pc} = \index$}
 \label{fig:compr_T6954_hcmh_Diff\index}
\end{subfigure}
}
\caption[]{
Dataset \data{Head1} compressed with \gls{PCA} (left column),
the optimized combination (middle) and their difference (right column). For
2 principal components \gls{PCA} suffers from reduced intensity
in the region of the lower jaw and exhibits more severe
streak artifacts in the neck region. For 4 principal components,
\gls{PCA} shows higher intensity at the right and top image border, but
without major impact on image quality. For 8 and 10 principal components,
the difference is negligible.
}
\label{fig:compr_T6954_hcmh}
\end{figure}

\begin{figure}[htbp]
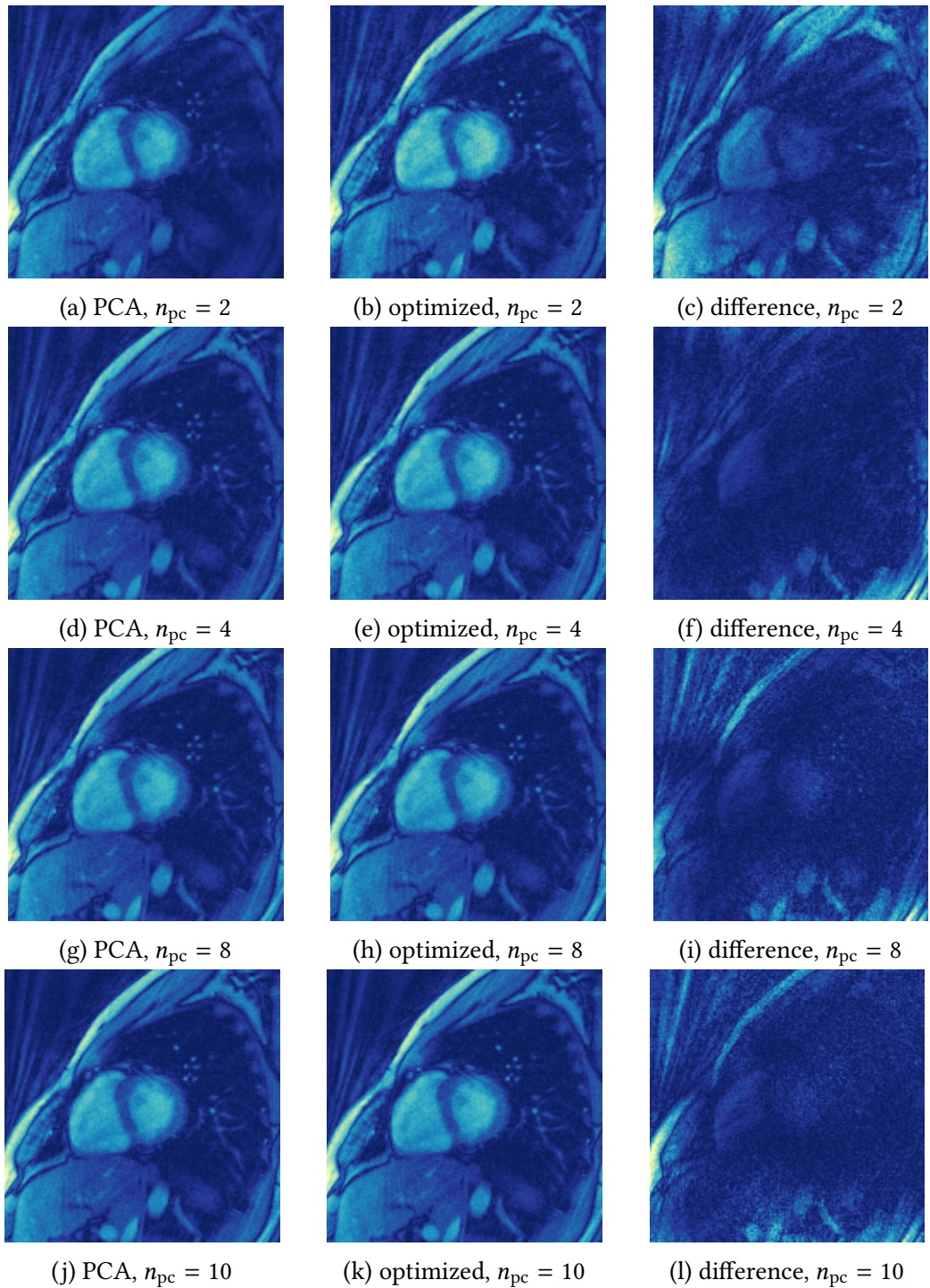

\centering
\foreach \index in {2,4,8,10} {%
\begin{subfigure}{0.3\textwidth}
  \centering
 \includegraphics[width=0.90\textwidth]
 {figures/matlab_figs/T18907_cardiac_hcmh/T18907_PCA_PD_\index.png}
 \caption{PCA, $n_\text{pc} = \index$}
 \label{fig:compr_T18907_cardiac_hcmh_PCA\index}
\end{subfigure}\hspace{0.01\textwidth}
\begin{subfigure}{0.3\textwidth}
  \centering
 \includegraphics[width=0.90\textwidth]
 {figures/matlab_figs/T18907_cardiac_hcmh/T18907_Soeren_PD_\index.png}
 \caption{optimized, $n_\text{pc} = \index$}
 \label{fig:compr_T18907_cardiac_hcmh_Soeren\index}
\end{subfigure}\hspace{0.01\textwidth}
\begin{subfigure}{0.3\textwidth}
  \centering
 \includegraphics[width=0.90\textwidth]
 {figures/matlab_figs/T18907_cardiac_hcmh/T18907_Diff_PD_\index.png}
 \caption{difference, $n_\text{pc} = \index$}
 \label{fig:compr_T18907_cardiac_hcmh_Diff\index}
\end{subfigure}
}
\caption[]{
Dataset \data{Heart1} (short-axis
view of the human heart) compressed with \gls{PCA} (left column),
the optimized combination (middle) and their difference (right column).
For 2 principal components, the optimized
combination has slightly higher signal in the region around the heart.
For 4, 8, and 10 principal components, the difference is negligible.
}
\label{fig:compr_T18907_cardiac_hcmh}
\end{figure}

\Crefrange{fig:compr_T16936_vr}{fig:compr_T18907_cardiac_hcmh}
provide an overview of the improvement that can be expected from using
the optimized compression algorithm over plain \gls{PCA}.

\begin{figure}[htbp]
\centering
\foreach \index in {2,4,8,10} {%
\begin{subfigure}{0.3\textwidth}
  \centering
 \includegraphics[width=0.90\textwidth]
 {figures/matlab_figs/T16936_vr/T16936_PCA_PD_\index.png}
 \caption{PCA, $n_\text{pc} = \index$}
 \label{fig:compr_T16936_vr_PCA\index}
\end{subfigure}\hspace{0.01\textwidth}
\begin{subfigure}{0.3\textwidth}
  \centering
 \includegraphics[width=0.90\textwidth]
 {figures/matlab_figs/T16936_vr/T16936_Soeren_PD_\index.png}
 \caption{optimized, $n_\text{pc} = \index$}
 \label{fig:compr_T16936_vr_Soeren\index}
\end{subfigure}\hspace{0.01\textwidth}
\begin{subfigure}{0.3\textwidth}
  \centering
 \includegraphics[width=0.90\textwidth]
 {figures/matlab_figs/T16936_vr/T16936_Diff_PD_\index.png}
 \caption{difference, $n_\text{pc} = \index$}
 \label{fig:compr_T16936_vr_Diff\index}
\end{subfigure}
}
\caption[]{
Dataset \data{Head2} compressed with \gls{PCA} (left column),
the optimized combination (middle) and their difference (right column).
For 2 und 4 principal
components
(\subref{fig:compr_T16936_vr_PCA2}--\subref{fig:compr_T16936_vr_Diff4})
the optimized combination yields clearly higher image quality, especially
regarding intensity distribution: while \subref{fig:compr_T16936_vr_PCA2}
contains almost no intensity in the region of the lower jaw,
\subref{fig:compr_T16936_vr_Soeren2} provides sufficient intensity there.
For 8 and 10 principal components, there is almost no visual difference.
}
\label{fig:compr_T16936_vr}
\end{figure}

As can be seen, for most datasets the improvement is limited to
low numbers of principal components (2 and 4); for 8 and 10 components,
the difference in all datasets is very small.

In the head datasets
(\cref{fig:compr_T16936_vr,fig:compr_T17086_brass_oct,fig:compr_T6954_hcmh})
for 2 principal components, the
improvement in image intensity uniformity can be cleary seen: The region
of the lower jaw shows very low intensity in the \gls{PCA} images and
only acceptable intensity in the optimized images. This effect is
less pronounced for higher numbers of principal components, where
\gls{PCA} has more uniform images also.

The optimized combination also reduces streak artifacts in some images,
most notably between
\cref{fig:compr_T16936_vr_PCA4,fig:compr_T16936_vr_Soeren4},
and between
\cref{fig:compr_T6954_hcmh_PCA2,fig:compr_T6954_hcmh_Soeren2}. This can
be understood as a consequence of image intensity uniformity: In \gls{PCA},
coils with very high but localized intensity will be preferred.
Streak
artifacts originate from these high intensity regions in these images.
By preferring uniform images, the relative intensity of the artifacts
is reduced compared to the rest of the image.

For the water phantom (\cref{fig:compr_T16677_phantom}) and the
cardiac data set (\cref{fig:compr_T18907_cardiac_hcmh}), the difference
is small for all numbers of principal components.

\FloatBarrier
\clearpage
\section{Discussion}
\label{sec:Compression_Discussion}
Comparing the optimized combination to the current \gls{PCA}
method shows
only improvement for 2 or 4 principal components. The same is true
for the possibility
of reducing the severity of streak artifacts.

The current implementation of the optimized combination requires
the proton density and coil profiles from a finished \gls{NLINV}. This is
a problem in real-time \gls{MRI}, where time is very constrained. It also
runs against the current design of \nlinvpp, requiring feedback
from the \texttt{reconstructor} to the \texttt{preprocessor}. This could
conceivably be overcome by approximating the proton density and
coil profiles: The proton density can be approximated as the
root-sum-of-squares of all individually gridded coil data
(which is the common method of combining multi-coil data,
see \textcite{RSS}), with the
coil profiles approximated as the individual gridding reconstruction
divided by the root-sum-of-squares image. But this only partly addresses the
problem, gridding and reconstructing data from 64 coils or more can still
take too much time to be feasible in real-time \gls{MRI}.

In the end, reducing the number of principal components to 8
provides too little benefit to justify the switch to the optimized method,
while reducing it to 4 or 6 principal components can lead to unacceptable
\gls{SNR} losses. 

\FloatBarrier
\cleardoublepage
\chapter{Coil Selection}
\label{sec:Selection}
\begin{figure}[tbp]
\centering
\begin{subfigure}{0.38\textwidth}
  \centering
 \includegraphics[width=0.95\textwidth]
 {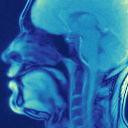}
 \caption{\data{Head1}}
 \label{fig:streak_Head1}
\end{subfigure}\hspace{0.1\textwidth}
\begin{subfigure}{0.38\textwidth}
  \centering
 \includegraphics[width=0.95\textwidth]
 {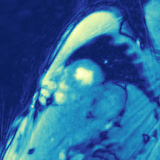}
 \caption{\data{Heart2}}
\label{fig:streak_Heart2}
\end{subfigure}
\vspace{1em}\\

\begin{subfigure}{0.38\textwidth}
  \centering
 \includegraphics[width=0.95\textwidth]
 {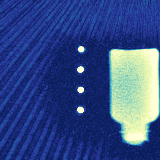}
 \caption{\data{Phan2}}
\label{fig:streak_Phan2}
\end{subfigure}\hspace{0.1\textwidth}
\begin{subfigure}{0.38\textwidth}
  \centering
 \includegraphics[width=0.95\textwidth]
 {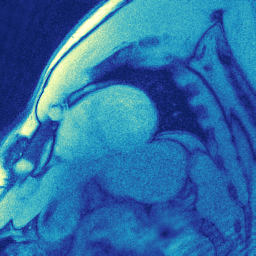}
 \caption{\data{Heart3}}
\label{fig:streak_Heart3}
\end{subfigure}
\vspace{1em}\\

\begin{subfigure}{0.38\textwidth}
  \centering
 \includegraphics[width=0.95\textwidth]
 {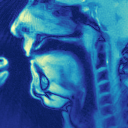}
 \caption{\data{Head4}}
\label{fig:streak_Head4}
\end{subfigure}\hspace{0.1\textwidth}
\begin{subfigure}{0.38\textwidth}
  \centering
 \includegraphics[width=0.95\textwidth]
 {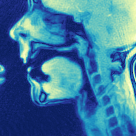}
 \caption{\data{Head5}}
\label{fig:streak_Head5}
\end{subfigure}

\caption[]{
Examples of streak artifacts in real-time images.
\subref{fig:streak_Head1} streaks in the neck region, lower right,
\subref{fig:streak_Heart2} streaks on the right side,
\subref{fig:streak_Phan2} streaks coming from top,
\subref{fig:streak_Heart3} streaks in lower right,
\subref{fig:streak_Head4} streaks in the neck region, lower
right,
\subref{fig:streak_Head5} streaks in middle right
(\data{Head5}). The streak artifacts are present even though
the median and non-local means postprocessing filters were
applied. In \subref{fig:streak_Phan2}, the streaks originate from
a counterflowing pipe outside of the \gls{FOV}.
}
\label{fig:streaks}
\end{figure}

Streak artifacts are a common problem in undersampled radial \gls{MRI}
\cites{streak1,streak2,streak3,streak4}.
Even for moderate undersampling image
quality can be compromised. Though real-time \gls{MRI} with
\gls{NLINV} does provide some mitigation (temporal regularization and
median filtering, for example), it is still a problem in a large number
of acquisitions.
\Cref{fig:streaks} show examples of \gls{NLINV} reconstructions
with both median and non-local means postprocessing
filters applied which still show streak artifacts.

The occurrence of streak artifacts can be understood as a confluence of
several factors: As \cref{sec:artifacts} showed, the degree of undersampling
necessary for real-time imaging will invariably lead to artifacts in
the resulting images. But most of these are unproblematic since the
mitigation techniques in \gls{NLINV} will greatly reduce the impact.
Often, problems arise from objects outside of the \gls{FOV}. Here, both the
static magnetic field and the gradient fields show larger inhomogeneities,
leading to overestimated and distorted signal intensity
\cite{streak_origin}, which in turn leads to more intense streak artifacts.
Streak artifacts are also often associated with fat which appears
hyperintense in the \gls{T1}-weighted \gls{FLASH} sequences used in
real-time \gls{MRI}, further amplifying the resulting artifacts.

\begin{figure}[tbp]
\centering
 \includegraphics[width=1.00\textwidth]
 {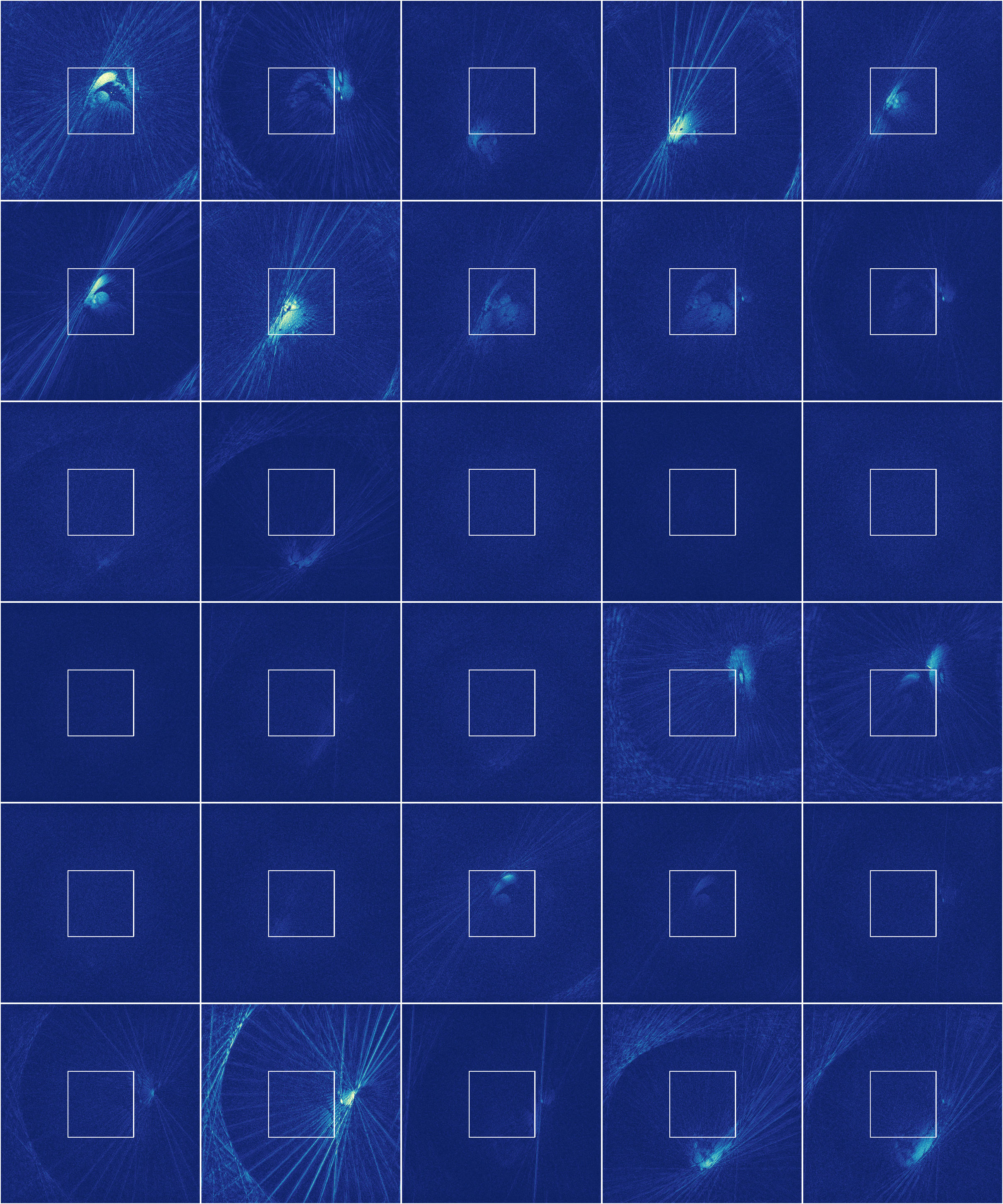}
\caption[]{
Gridding reconstructions of all 30 coils for \data{Heart2}. The white
square indicates the selected \gls{FOV}. The three times
larger size of the image matrix is a
consequence of 2 times oversampling and $1.5$ times
overgridding (see \cref{sec:data}). Not all coils
contribute useful signal inside the \gls{FOV}: Some contain very little
signal at all, while some only contain signal outside of the \gls{FOV}.
All images are shown with the same absolute windowing.
}
\label{fig:individual}
\end{figure}

\begin{figure}[tbp]
\centering
\begin{subfigure}{0.38\textwidth}
  \centering
 \includegraphics[width=0.95\textwidth]
 {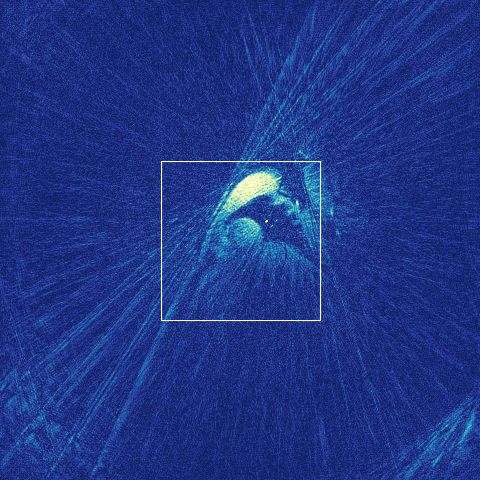}
 \caption{}
 \label{fig:individual_useful}
\end{subfigure}\hspace{0.1\textwidth}
\begin{subfigure}{0.38\textwidth}
  \centering
 \includegraphics[width=0.95\textwidth]
 {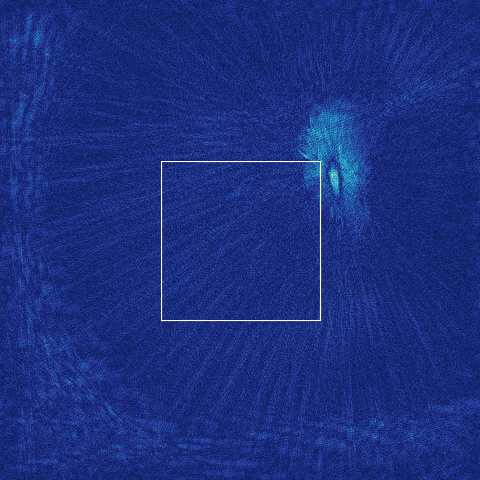}
 \caption{}
 \label{fig:individual_outside}
\end{subfigure}
\vspace{1em}\\

\begin{subfigure}{0.38\textwidth}
  \centering
 \includegraphics[width=0.95\textwidth]
 {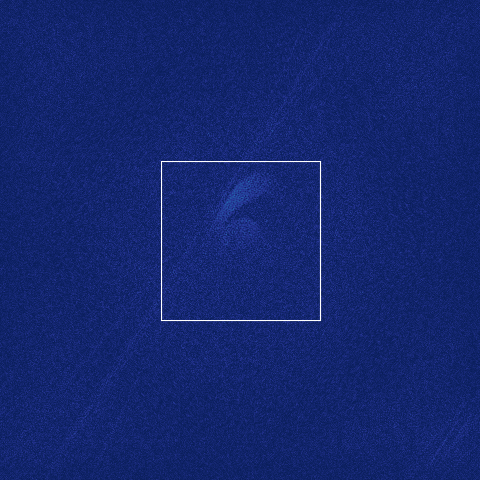}
 \caption{}
 \label{fig:individual_empty}
\end{subfigure}\hspace{0.1\textwidth}
\begin{subfigure}{0.38\textwidth}
  \centering
 \includegraphics[width=0.95\textwidth]
 {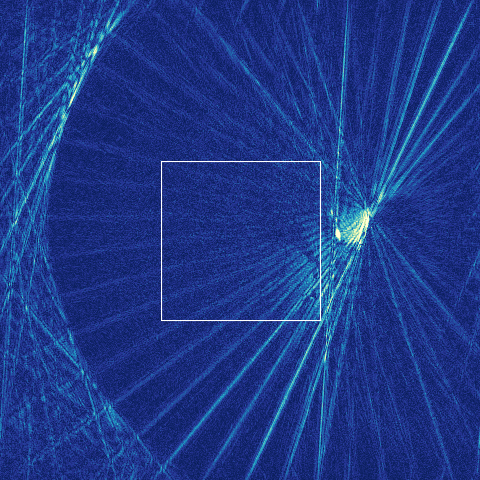}
 \caption{}
 \label{fig:individual_streak}
\end{subfigure}

\caption[]{
Examples of gridding reconstructions of individual coils for
\data{Heart2}, showing a
subset of \cref{fig:individual}. \subref{fig:individual_useful} shows a coil
which, while containing some streaks, contains useful signal inside
the \gls{FOV}. \subref{fig:individual_outside} contains signal mostly
outside the \gls{FOV}. \subref{fig:individual_empty} contains almost no signal,
indicating a coil far from the imaged plane. \subref{fig:individual_streak}
contains unwanted streak artifacts intersecting the \gls{FOV}.
All images are shown with the same absolute windowing.
}
\label{fig:individual2}
\end{figure}

\begin{figure}[tbp]
\centering
\foreach \index in {all,man} {%
\begin{subfigure}{0.38\textwidth}
  \centering
 \includegraphics[width=0.95\textwidth]
 {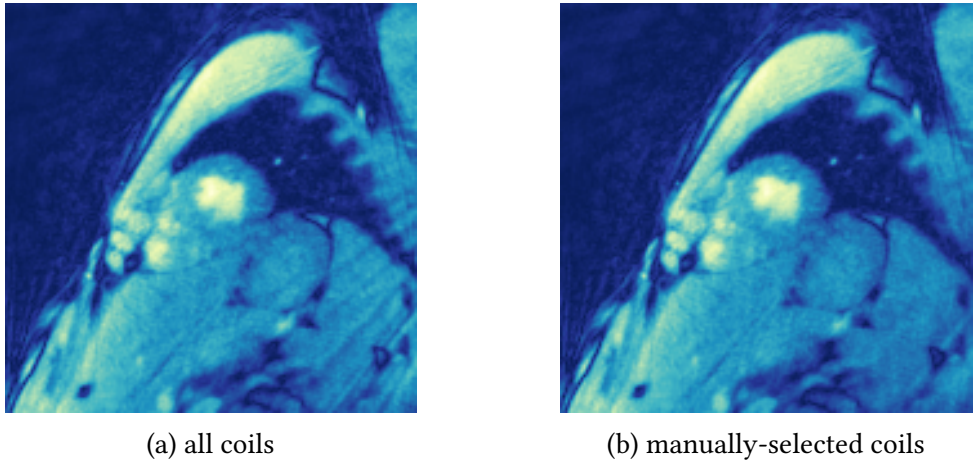}
 \caption{\repr{\index} coils}
\label{fig:man_comparison_\index}
\end{subfigure}\hspace{0.1\textwidth}
}\hspace{-0.1\textwidth}

\caption[]{
Comparison of \gls{NLINV} reconstruction of \data{Heart2} with
\subref{fig:man_comparison_all} all coils and
\subref{fig:man_comparison_man} without the coil shown in
\cref{fig:individual_streak}. As can be seen, the streaks in the middle right
have been removed
almost completely by excluding a single coil from the reconstruction.
}
\label{fig:man_comparison}
\end{figure}

The central insight justifying coil selection for streak artifact
reduction is that only some coils contain data leading to
streak artifacts in the reconstructed image:
This can be seen by studying individual coil
images. These can be reconstructed with plain gridding and
subsequent Fourier-transform of individual coil data. Some examples
of such individual coil images are shown in
\cref{fig:individual,fig:individual2}, with prominent streak
artifacts in only a subset of coils.
Furthermore, excluding those coils during reconstruction can greatly
reduce or even
remove streak artifacts. This can be seen in \cref{fig:man_comparison},
which shows the comparison of the reconstruction of the dataset shown in
\cref{fig:individual} with all coils (\cref{fig:man_comparison_all})
and with one coil removed (\cref{fig:man_comparison_man}).
Coil selection has to be applied
before coil compression: After coil compression, individual coils cannot
be separated out anymore, having been intermixed into
the set of virtual coils.

\FloatBarrier
\section{Current Methods}
\label{sec:Selection_methods}

\subsection{Manual Selection}
\label{sec:Selection_man}
The most simple way of coil selection is manual selection, that means turning
off unwanted coils before a measurement or ignoring some coils during
reconstruction. For almost all \gls{MRI} acquisitions, some coils are
turned off: For example for a head measurement, spine coils integrated in
the patient table are generally turned off. However, this is not feasible for
routine use, since it is in general not possible to know which coils
will produce artifacts.

\subsection{Xue et al.}
\label{sec:Selection_xue}

In \cite{Xue}, \citeauthor{Xue} introduced a method which seeks to
provide automatic coil selection.
It is based on the insight that streak artifacts in radial
imaging are associated with outer $k$-space: Low resolution images
will typically not contain
excessive streak artifacts. This can be immediately understood as a
consequence of sampling: Close to the $k$-space center, there is inherently
closer sampling in radial acquisitions: For a fixed number of spokes,
the violation of the Nyquist criterion is stronger further out in $k$-space.
The algorithm is:

\begin{enumerate}
 \item Grid and reconstruct the magnitude images $I_\text{orig}$ for all coils.
 \item Apply a low-pass Hanning filter to the $k$-space data and reconstruct
  low resolution magnitude images $I_\text{ref}$.
 \item Calculate the streak ratio $R_\text{streak}$ given by
   $R_\text{streak} = \frac{\mean(\abs{I_\text{orig} - I_\text{ref}})}
                            {\mean(I_\text{ref})}$.
  \item Exclude coils where $R_\text{streak}$ is larger than a predetermined
    threshold.
\end{enumerate}
However, this approach is not fully automatic, it needs a predefined
threshold for its streak ratio. This threshold needs to be manually found
for the current application. Furthermore, it relies on image-space data,
which means that data needs to be gridded and Fourier-transformed. Especially
gridding is a very expensive operation, limiting the usefulness of this
approach for real-time \gls{MRI}. For these reasons, it was neither implemented
nor evaluated in this thesis.

\subsection{Grimm et al.}
\label{sec:Selection_grimm}
\textcite{TB} introduced an alternative coil selection method based
on the algorithm of \textcite{Xue} which tries
to overcome these deficiencies. It replaces the threshold above with
a clustering method, thereby eliminating the need to manually set a
threshold. Furthermore, it uses $k$-space data directly and does not need
gridding.

The algorithm is:
\begin{enumerate}
  \item Generate a high-pass filtered variant $h_n$ and a low-pass filtered
    variant $l_n$ of the ungridded $k$-space data for each coil $n$.
  \item Calculate the streak ratio $\tilde{R}_n$ by
    $\tilde{R}_n = \frac{\norm{h_n}_2}{\norm{l_n}_2}$.
  \item Apply k-means clustering to separate the coils into two groups
    based on their $\tilde{R}_n$. Two values of $\tilde{R}_n$ are randomly
    chosen as initial cluster centers. \label{step:TB_kmeans}
  \item Calculate the distance between the clustering centers. If it is less
    than twice the average standard deviation,
    repeat step \ref{step:TB_kmeans} up to $n_\text{tries}$ times.
    Otherwise,
    exclude the group of coils with high $\tilde{R}_n$. \label{step:TB_safety}
\end{enumerate}
Apart from the clustering and use of $k$-space data, it
changes the Hamming window to a simple box window separating the data
into high and low frequency parts, and changes the mean to the $L^2$-norm.
Furthermore, the k-means algorithm can automatically find appropriate groups,
while also providing a check to not exclude coils
if no sufficiently separated
groups can be found\footnote{This is necessary since k-means will almost
always generate two groups (if initialized with two centers),
regardless of whether those groups exist in the data or not. Without such a
basic check, the algorithm might exclude coils even if none of them
contain any artifacts.}.
Furthermore, since it
uses the ungridded raw data, it is very fast. The repetition of
k-means\footnote{Up to $n_\text{tries} = 100$ times.}
is done to ensure that the random choice of the initial cluster centers does
not prevent k-means from excluding coils. The k-means algorithm itself and its
random initialization make the algorithm non-deterministic: It is not
guaranteed that consecutive runs will exclude the same coils. But for most
datasets (and for all datasets shown in this thesis), the algorithm
consistently gives the same results. Since no generally
applicable value for the cutoff between high and low $k$-space was given,
several cutoffs were tried. In the end, for an acquisition of
$n_\text{samples}$ samples, the inner $\sfrac{n_\text{samples}}{4}$ were
defined as the low $k$-space part $l_n$, with the rest assigned to $h_n$.

\begin{figure}[tbp]
\centering
\foreach \index in {all,TB} {%
\begin{subfigure}{0.38\textwidth}
  \centering
 \includegraphics[width=0.95\textwidth]
 {figures/matlab_figs/T6954_hcmh/T6954_nlinvpp_\index_nofilter_crop_t3.png}
 \caption{\data{Head1}, \repr{\index} coils}
 \label{fig:TB_comparison_Head1_\index}
\end{subfigure}\hspace{0.1\textwidth}
}\hspace{-0.1\textwidth}\vspace{1em}\\

\foreach \index in {all,TB} {%
\begin{subfigure}{0.38\textwidth}
  \centering
 \includegraphics[width=0.95\textwidth]
 {figures/matlab_figs/T11500_perfusion_pha/T11500_nlinvpp_\index_nofilter_crop_t3.png}
 \caption{\data{Phan2}, \repr{\index} coils}
  \label{fig:TB_comparison_Phan2_\index}
\end{subfigure}\hspace{0.1\textwidth}
}\hspace{-0.1\textwidth}\vspace{1em}\\

\foreach \index in {all,TB} {%
\begin{subfigure}{0.38\textwidth}
  \centering
 \includegraphics[width=0.95\textwidth]
 {figures/matlab_figs/T10366_cardiac_multislice/T10366_nlinvpp_\index_nofilter_crop_t3.png}
 \caption{\data{Heart2}, \repr{\index} coils}
  \label{fig:TB_comparison_Heart2_\index}
\end{subfigure}\hspace{0.1\textwidth}
}\hspace{-0.1\textwidth}

\caption[]{Comparison between using all coils (left column)
and \posscite{TB} selection algorithm (right column) for
several datasets.
For \data{Head1}
(\subref{fig:TB_comparison_Head1_all}--\subref{fig:TB_comparison_Head1_TB}),
the streak artifacts in the neck region are removed almost completely,
while the other datasets still contain significant artifacts. In all
examples, coils were removed; the criterion described in
step \ref{step:TB_safety} did not apply.
}
\label{fig:TB_comparison}
\end{figure}

\begin{figure}[tbp]
\centering
\begin{subfigure}{0.80\textwidth}
  \centering
 \includegraphics[width=0.95\textwidth]
 {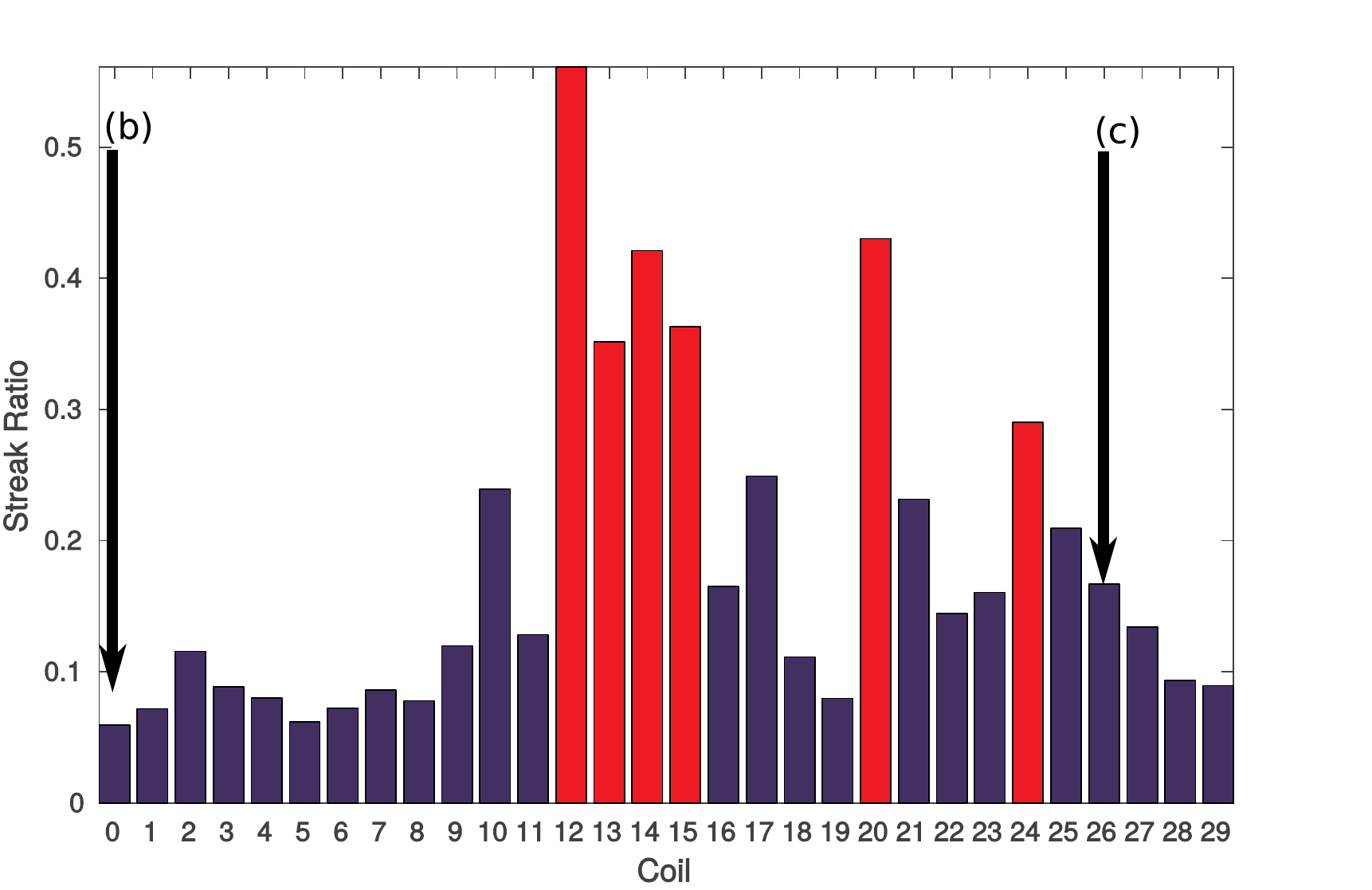}
 \caption{}
 \label{fig:TB_streak_ratios_bar1}
\end{subfigure}
\vspace{1em}\\
\begin{subfigure}{0.38\textwidth}
  \centering
 \includegraphics[width=0.95\textwidth]
 {figures/matlab_figs/T10366_cardiac_multislice/coil00-gridding_reco.png}
 \caption{}
 \label{fig:T10366_coil00}
\end{subfigure}\hspace{0.1\textwidth}
\begin{subfigure}{0.38\textwidth}
  \centering
 \includegraphics[width=0.95\textwidth]
 {figures/matlab_figs/T10366_cardiac_multislice/coil26-gridding_reco.png}
 \caption{}
 \label{fig:T10366_coil26}
\end{subfigure}

\caption[]{
Example of calculated streak ratios of \posscite[s][method]{TB} for
\data{Heart2}.
\subref{fig:TB_streak_ratios_bar1} shows the calculated streak ratios;
gridding reconstructions of the indicated coils are shown
in \subref{fig:T10366_coil00}
and \subref{fig:T10366_coil26}. $\tilde{R_n}$
is not a good measure for the streak content in the individual coils,
since \subref{fig:T10366_coil26} is the single coil responsible
for streak artifacts in the final reconstruction
(see \cref{fig:man_comparison}).
}
\label{fig:TB_streak_ratios}
\end{figure}

\begin{figure}[tbp]
\centering
\begin{subfigure}{0.80\textwidth}
  \centering
  \includegraphics[width=0.95\textwidth]
 {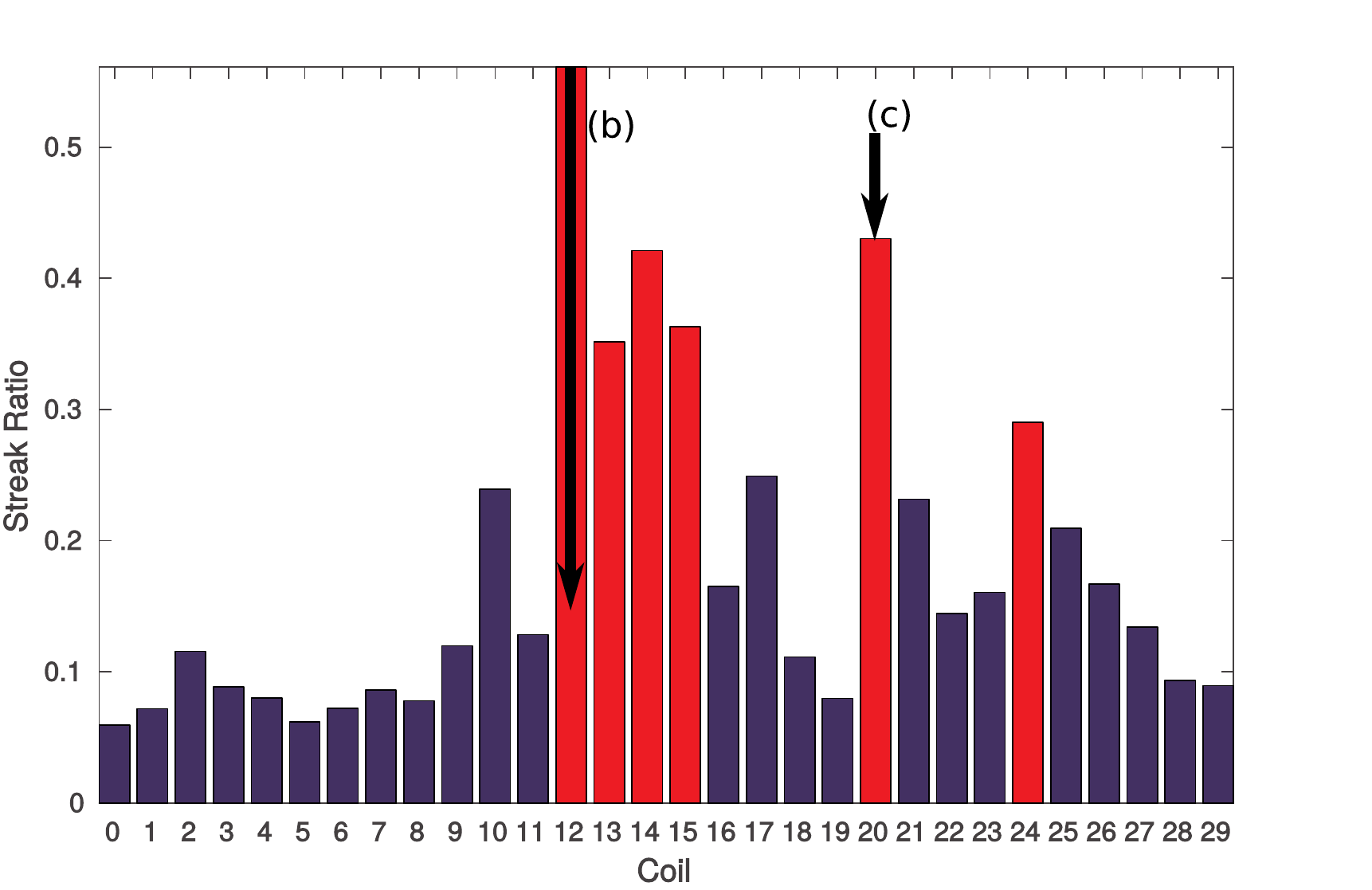}
 \caption{}
 \label{fig:TB_streak_ratios_bar2}
\end{subfigure}
\vspace{1em}\\

\begin{subfigure}{0.38\textwidth}
  \centering
 \includegraphics[width=0.95\textwidth]
 {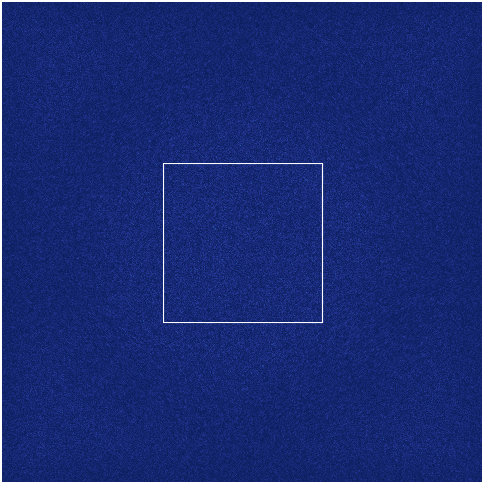}
 \caption{}
\label{fig:T10366_coil12}
\end{subfigure}\hspace{0.1\textwidth}
\begin{subfigure}{0.38\textwidth}
  \centering
 \includegraphics[width=0.95\textwidth]
 {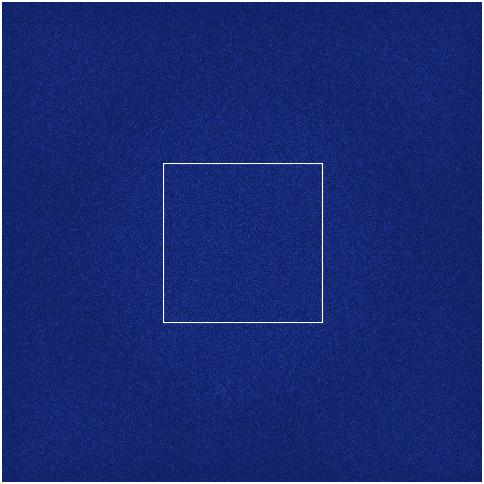}
 \caption{}
\label{fig:T10366_coil20}
\end{subfigure}
\caption[]{
Example of calculated streak ratios of \posscite[s][method]{TB} for
\data{Heart2}.
\subref{fig:TB_streak_ratios_bar2} shows the calculated streak ratios,
gridding reconstructions of the indicated coils are shown
in \subref{fig:T10366_coil12}
and \subref{fig:T10366_coil20}. High streak ratios were
assigned to both coils, while neither contains useful signal.
}
\label{fig:TB_empty_coils}
\end{figure}

\Cref{fig:TB_comparison} shows a comparison between using all coils and
using \posscite{TB} selection for some datasets. As can be seen,
\posscite[s][method]{TB} does not remove the prominent
streak artifacts in a number of acquisitions. By examining the streak
ratios $\tilde{R}_n$ assigned to various coils
(see \cref{fig:TB_streak_ratios}), it can be seen that often the
values of $\tilde{R}_n$ do not reflect the severity of the artifacts.
Another problem is the algorithm's behavior in the presence of mostly empty
coils: Even if a coil does not contain usable signal, white receiver noise will
still be present, possibly dominating the calculated streak ratios.
This effect can be seen in \cref{fig:TB_empty_coils}, for example.
While excluding these coils does not negatively influence image quality,
it also does not remove streak artifacts.
In conclusion, the performance for this method for real-time \gls{MRI} was not
satisfactory.

\FloatBarrier
\section{Sinogram-based Selection}
\label{sec:Selection_sino}
In order to address the shortcomings of \posscite[s][method]{TB}
in the
context of real-time \gls{MRI}, a modified algorithm was developed on its
basis.
The general idea is to use sinograms to reach a compromise between the full
image space information needed in \posscite{Xue} algorithm and
the pure $k$-space implementation of \textcite{TB}. Furthermore, low
intensity coils are identified and ignored so that they do not
influence the streak ratio calculation. They are, however, not
excluded, since the subsequent coil compression will weigh them
appropriately.

\begin{figure}[tb]
\centering
\begin{subfigure}{0.80\textwidth}
\centering
 \includegraphics[width=0.95\textwidth]
 {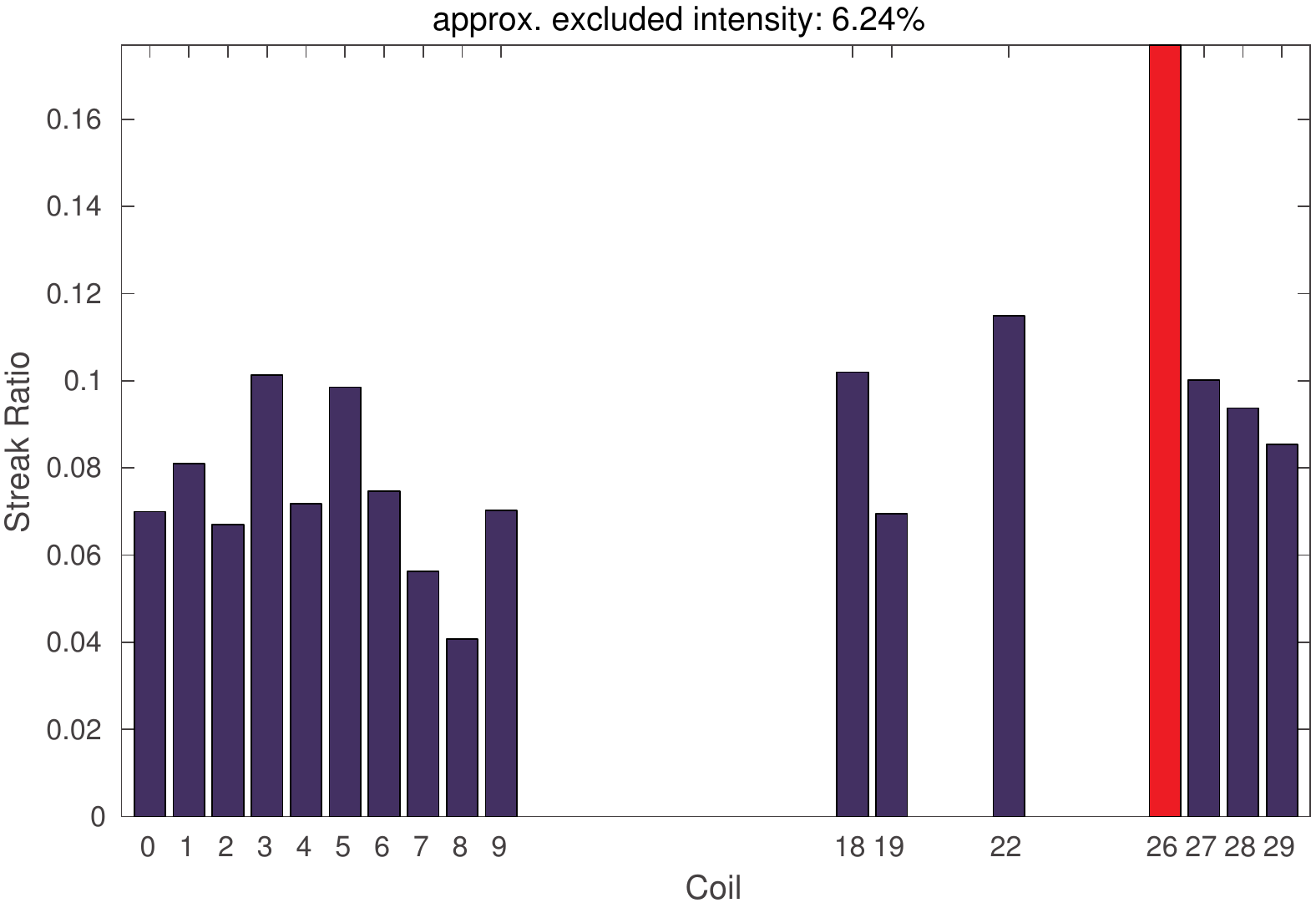}
  \caption[]{}
  \label{fig:sino_streak_ratios_bar}
\end{subfigure}
\\\vspace{1em}
\begin{subfigure}{0.80\textwidth}
\centering
 \includegraphics[width=0.95\textwidth]
 {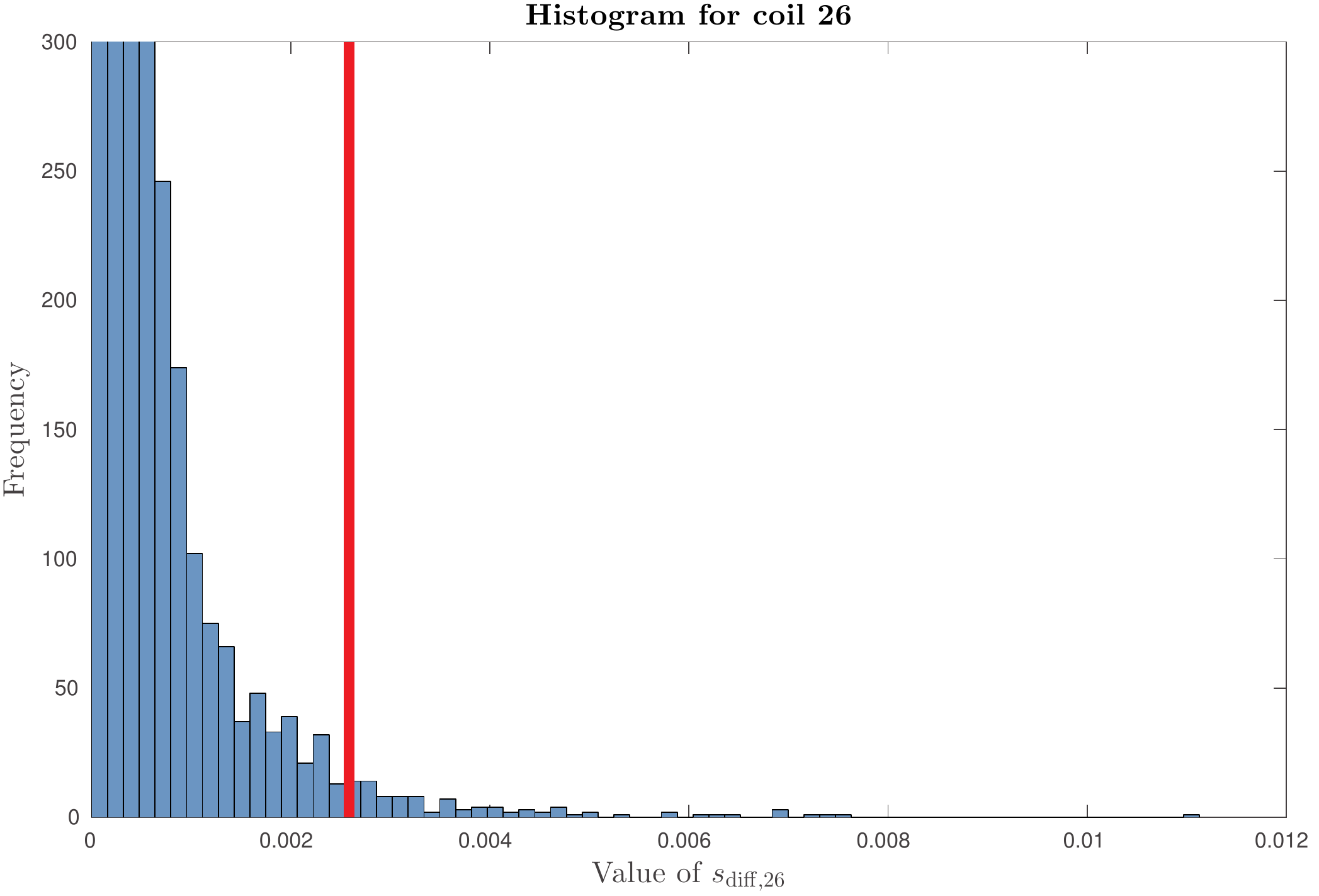}
 \caption[]{}
 \label{fig:histogram}
 \end{subfigure}
\caption[]{
\subref{fig:sino_streak_ratios_bar}: Example of the calculated
streak ratios using the sinogram-based seletion for dataset
\data{Heart2}. The coils
without any streak ratio were ignored in step \ref{step:empty}.
The coil indicated in red is to be removed, which is the same coil
that was removed as part of the manual selection shown in
\cref{fig:man_comparison}.
\subref{fig:histogram}: Histogram of $s_{\text{diff},n}$
for the indicated coil in \subref{fig:sino_streak_ratios_bar}
(see \cref{fig:individual_streak} for a gridding reconstruction). The frequency axis is truncated to
emphasize the small frequency values for larger values of
$s_{\text{diff},n}$. The red vertical line indicates the threshold
calculated in step \ref{step:thresholding}. The value of the threshold
is calculated so that the majority of the noisy low value pixels
are excluded.
}
\label{fig:sino_method}
\end{figure}

\begin{figure}[tb]
\centering
\begin{subfigure}{0.44\textwidth}
\centering
 \includegraphics[width=0.95\textwidth]
 {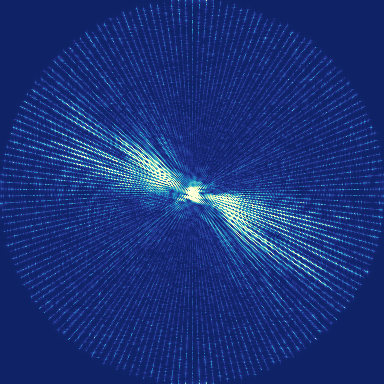}
  \caption[]{}
  \label{fig:sino_coil33_kspace}
\end{subfigure}\hspace{0.1\textwidth}
\begin{subfigure}{0.44\textwidth}
\centering
 \includegraphics[width=0.95\textwidth]
 {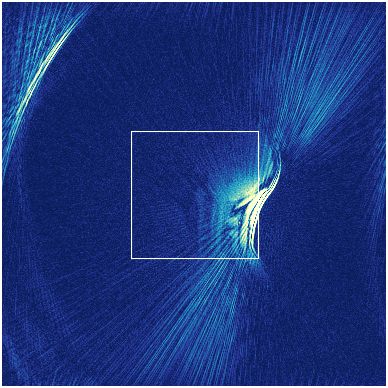}
  \caption[]{}
  \label{fig:sino_coil33}
\end{subfigure}\\\vspace{1em}
\begin{subfigure}{0.28\textwidth}
\centering
 \includegraphics[width=0.60\textwidth]
 {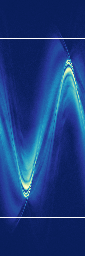}
  \caption[]{}
  \label{fig:sino_S}
\end{subfigure}\hspace{0.05\textwidth}
\begin{subfigure}{0.28\textwidth}
\centering
 \includegraphics[width=0.60\textwidth]
 {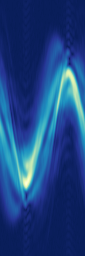}
 \caption[]{}
 \label{fig:sino_Slow}
 \end{subfigure}\hspace{0.05\textwidth}
\begin{subfigure}{0.28\textwidth}
\centering
 \includegraphics[width=0.60\textwidth]
  {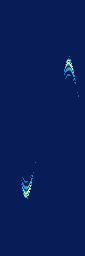}
 \caption[]{}
 \label{fig:sino_Sdiff}
 \end{subfigure}
\caption[]{
\subref{fig:sino_coil33_kspace} Gridded $k$-space and
\subref{fig:sino_coil33} gridding reconstruction of
the first full frame of a streaked coil
removed by sinogram-based selection
from dataset \data{Head1} (see \cref{fig:sino_comparison_Head1_all} for
an \gls{NLINV} reconstruction).
\subref{fig:sino_S} High- and \subref{fig:sino_Slow} low-resolution sinogram,
\subref{fig:sino_Sdiff} thresholded difference of sinograms
for the same coil. The spatial location of the streak origin
can be clearly identified in the thresholded difference of the
low- and high-resolution sinograms, while it cannot be identified
in the gridded $k$-space in \subref{fig:sino_coil33_kspace}.
In \subref{fig:sino_S}, the area used to estimate the signal contribution
to the \gls{FOV} is indicated in white (see step \ref{step:FOV_contrib}).
}
\label{fig:sino_sinograms}
\end{figure}

The algorithm, termed sinogram-based selection, is as follows:
\begin{enumerate}
  \item For each coil $n$, generate a high-pass filtered variant
    $h_n$ and a low-pass filtered
    variant $l_n$ of the ungridded $k$-space data.
  \item Calculate the 1D \gls{DFT} along the sample direction, yielding
    high and low resolution sinograms $s_{h,n}$ and $s_{l,n}$.
  \item Approximate the
    signal contribution $F_n$ of each coil to the \gls{FOV} by the $L^2$ norm
    of the inner part of the high resolution sinogram. The width of this
    inner part is the length of the diagonal of the
    \gls{FOV}. Normalize by dividing by $\sum_nF_n$. \label{step:FOV_contrib}
  \item Calculate the mean $\mu_F$ and the standard deviation $\sigma_F$
    of the \gls{FOV} contribution $F_n$. Any coil with
    $F_n < \frac{1}{3}(\mu_F + \sigma_F)$ is ignored for the rest of the
    algorithm. Renormalize $F_n$ for each coil. \label{step:empty}
  \item Calculate the magnitude of the complex difference
    $s_{\text{diff},n} = \abs{s_{h,n} - s_{l,n}}$, its standard deviation
    $\sigma_{\text{diff},n}$ and its mean $\mu_{\text{diff},n}$.
  \item For each coil $n$, apply thresholding to $s_{\text{diff},n}$ with
    threshold $T = \mu_{\text{diff},n} + 4\cdot\sigma_{\text{diff},n}$,
    that means set $s_{\text{diff},n}$ to zero where it is smaller than $T$,
    generating the thresholded $s_{\text{diff},n}^T$. \label{step:thresholding}
  \item Calculate the streak ratio $R_n$ for each coil as
    $R_n = \frac{\norm{s_{\text{diff},n}^T}}{\norm{s_{l,n}}}$.
  \item Apply k-means to sort the streak ratios into two groups clustered
    around two centers. \label{step:kmeans}
  \item If the ratio between the high and low centers is less than 2,
    repeat step \ref{step:kmeans} up to $100$ times.
  \item If the combined \gls{FOV} contribution of the coils  in the
    high group is more than \SI{20}{\percent}, exclude the coils with
    the highest streak ratio until \SI{20}{\percent} excluded intensity
    is reached, otherwise exclude all coils in the high group.
\end{enumerate}
The thresholding of the difference sinograms is done as
a crude form of noise suppression: By studying the histograms
(see \cref{fig:histogram} as an example), it can be seen that low
intensity pixels dominate the difference sinogram. Thresholding in the
way described in step \ref{step:thresholding} can filter these pixels and
thereby emphasize the important difference between the sinograms
contained in few pixels. This is also the reason why sinograms
are used: as shown in \cref{fig:sino_sinograms}, the spatial origin
of streak artifacts can be clearly determined from the difference
sinogram. In $k$-space, this difference is spread out over the
entire space, so no useful thresholding can be done.

The center ratio is used as a criterion for deciding if the  k-means
algorithm
managed to identify separate groups; if the center ratio is small the distance
between the groups is also small.
The estimation of the contribution to the \gls{FOV} is also used as a safety
measure: sufficient signal in the resulting images has to be guaranteed,
even at the expense of remaining artifacts. \SI{20}{\percent} has been
used as compromise between allowing the algorithm freedom to exclude coils
and the need to preserve signal content in the reconstructed images. It is,
however, an estimate only, since the contribution of each coil is modified
by channel compression and by \gls{NLINV} itself. Since
sinogram-based selection shares the same k-means
initialization as \posscite[s][method]{TB}, it is also not guaranteed to
be deterministic.

\Cref{fig:sino_comparison_good} shows a comparison between
using all coils and the proposed selection for datasets where it
worked well: comparing \cref{fig:sino_comparison_good}
to \cref{fig:TB_comparison} shows marked improvement over
\posscite{TB} selection for real-time data.

But problems remain for other datasets:
In \cref{fig:sino_bad_vr_all,fig:sino_bad_vr_integr}, the streak artifacts
in the neck region are greatly reduced, but the nose region still contains
significant artifacts. Similar behaviour can be seen
in \cref{fig:sino_bad_sdohr_all,fig:sino_bad_sdohr_integr}: Here, too,
streak artifacts in the nose region are unaffected, while the artifacts in
the neck region are only slightly reduced. Bar diagrams of the streak ratios
for these datasets can be found in \cref{sec:figures}.

\begin{figure}[tbp]
\centering
\foreach \index in {all,TB,integr} {%
\begin{subfigure}{0.30\textwidth}
  \centering
 \includegraphics[width=0.95\textwidth]
 {figures/matlab_figs/T6954_hcmh/T6954_nlinvpp_\index_nofilter_crop_t3.png}
 \caption{\repr{\index} coils}
\label{fig:sino_comparison_Head1_\index}
\end{subfigure}\hspace{0.01\textwidth}
}\hspace{-0.01\textwidth}\vspace{1em}\\

\foreach \index in {all,TB,integr} {%
\begin{subfigure}{0.30\textwidth}
  \centering
 \includegraphics[width=0.95\textwidth]
 {figures/matlab_figs/T11500_perfusion_pha/T11500_nlinvpp_\index_nofilter_crop_t3.png}
 \caption{\repr{\index} coils}
\label{fig:sino_comparison_Phan2_\index}
\end{subfigure}\hspace{0.01\textwidth}
}\hspace{-0.01\textwidth}\vspace{1em}\\

\foreach \index in {all,TB,integr} {%
\begin{subfigure}{0.30\textwidth}
  \centering
 \includegraphics[width=0.95\textwidth]
 {figures/matlab_figs/T10366_cardiac_multislice/T10366_nlinvpp_\index_nofilter_crop_t3.png}
 \caption{\repr{\index} coils}
\label{fig:sino_comparison_Heart2_\index}
\end{subfigure}\hspace{0.01\textwidth}
}\hspace{-0.01\textwidth}

\caption[]{
Comparison between using all coils (left column),
\posscite[s][method]{TB} (center column) and the
proposed sinogram-based selection (right column) for several datasets.
\subref{fig:sino_comparison_Head1_all} (\data{Head1}) shows streak artifacts in the
neck region, which are greatly reduced in
\subref{fig:sino_comparison_Head1_integr}, as are the prominent
streak artifacts in \subref{fig:sino_comparison_Phan2_all} (\data{Phan2})
and \subref{fig:sino_comparison_Heart2_all} (\data{Heart2}).
The artifacts in \subref{fig:sino_comparison_Phan2_all} and
\subref{fig:sino_comparison_Heart2_all}
are not removed by \posscite[s][method]{TB}.
}
\label{fig:sino_comparison_good}
\end{figure}

\begin{figure}[tbp]
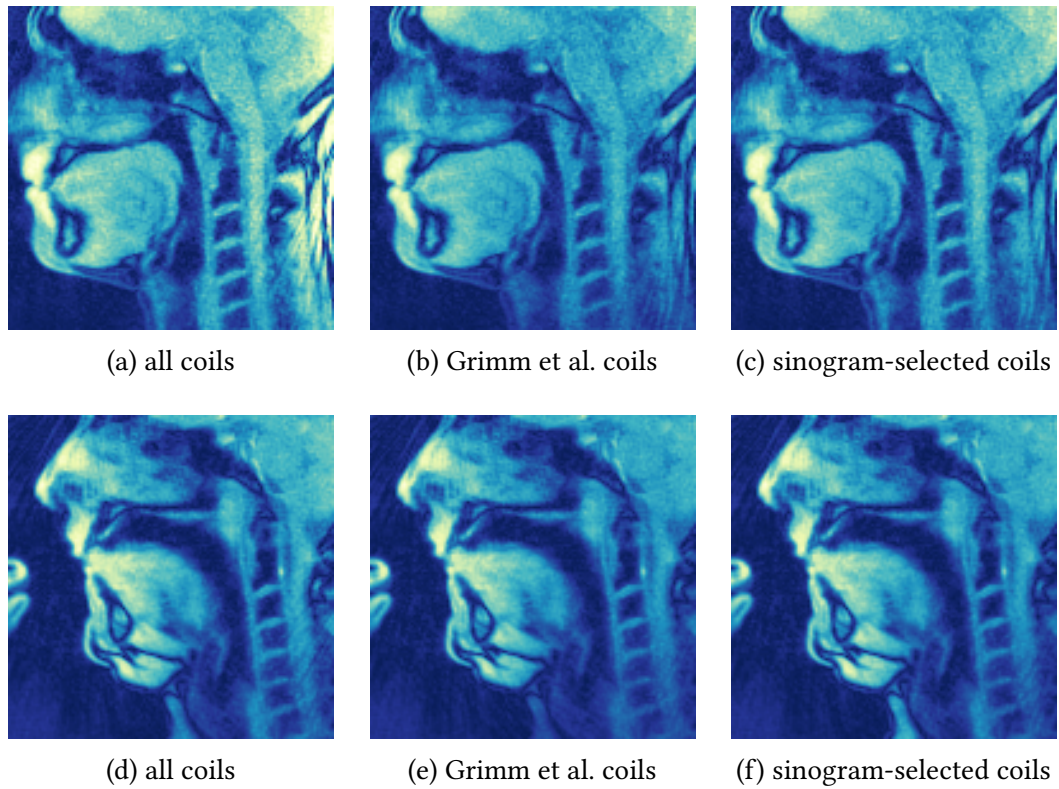

\centering
\foreach \index in {all,TB,integr} {%
\begin{subfigure}{0.30\textwidth}
  \centering
 \includegraphics[width=0.95\textwidth]
 {figures/matlab_figs/T16936_vr/T16936_nlinvpp_\index_nofilter_crop_t3.png}
 \caption{\repr{\index} coils}
 \label{fig:sino_bad_vr_\index}
\end{subfigure}\hspace{0.01\textwidth}
}\hspace{-0.01\textwidth}\vspace{1em}\\

\foreach \index in {all,TB,integr} {%
\begin{subfigure}{0.30\textwidth}
  \centering
 \includegraphics[width=0.95\textwidth]
 {figures/matlab_figs/T25921_sdohr/T25921_nlinvpp_\index_nofilter_crop_t3.png}
 \caption{\repr{\index} coils}
  \label{fig:sino_bad_sdohr_\index}
\end{subfigure}\hspace{0.01\textwidth}
}\hspace{-0.01\textwidth}\vspace{1em}\\


\caption[]{
Comparison between using all coil (left column), \posscite[s][method]{TB}
(center column), and
the proposed sinogram-based selection (right column) for additional
datasets.
While both \subref{fig:sino_bad_vr_TB} and
\subref{fig:sino_bad_vr_integr} (\data{Head2}) show reduced
streak artifacts in the
neck region, some artifacts are still present there. Furthermore,
strong artifacts in the nose region remain unchanged compared to
\subref{fig:sino_bad_vr_all}.
Similarly,  both \subref{fig:sino_bad_sdohr_TB} and
\subref{fig:sino_bad_sdohr_integr} (\data{Head4})
show artifacts in the
nose region of unchanged strength and only slightly reduced artifacts in
the neck region compared to \subref{fig:sino_bad_sdohr_all}.
}
\label{fig:sino_comparison_bad}
\end{figure}

Because of the improved image quality gained by sinogram-based
selection, a \texttt{C++} implementation was written and added to
the \texttt{preprocessor} of \nlinvpp as part of this thesis.
This enables online use directly on the \gls{MRI} scanner.

This implementation uses the existing
calibration data\footnote{The raw data of the first full frame.}
which is already used to calculate the gradient delay correction
values and the
coil compression matrix (\gls{PCA}).
Coil selection is now a step before these two others, deleting
coils from the calibration data if necessary. Gradient delay values and
\gls{PCA} matrix are then calculated on this filtered
calibration data. But the raw data sent on by the \texttt{datasource}
still contains all coils, so it is incompatible with the calculated
\gls{PCA} matrix. To overcome this, columns of all zeros
are added to the \gls{PCA} matrix. Since the
matrix-matrix product between incoming raw data and the \gls{PCA} matrix
has to be calculated regardless of whether coil selection is used or not,
this makes applying the coil selection to subsequent frames free in
terms of added calculation time.
So coil selection only has a cost during initial calibration of the
\texttt{preprocessor}.
This cost is, however, quite small in terms of wall-clock time:
For a common dataset with 64 coils the sinogram-based selection
takes less than \SI{0.5}{\second}.
Furthermore, the overall effect is that
excluding coils through coil selection is indistinguishable from
not having measured these coils at all.

\FloatBarrier
\clearpage
\section{Discussion}
\label{sec:Selection_Discussion}
The novel sinogram-based selection algorithm
has been shown to provide better
artifact-reduction than previous algorithms in the context of
real-time \gls{MRI}. But nontrivial problems with some datasets
remain, where the streak artifact reduction is not totally
satisfactory.

As \posscite[s][method]{TB}, the proposed sinogram-based
selection is also non-deterministic.
While stable for most datasets, some datasets will
show different selections on consecutive runs, leading to non-reproducible
reconstructions. A further problem is the non-determined runtime: Since
the k-means algorithm will try multiple times to find
a selection matching all
criteria, the number of tries can vary from run to run. Incidentally, the
worst runtime is for datasets where no coils will be excluded: Here the
maximum number of tries is exhausted, leading to longer runtimes which
are still below \SI{1}{\second}.

In general, since the streak ratios used in both
algorithms are scalars, the k-means algorithm acts as a simple threshold.
Finding an appropriate threshold with which the k-means algorithm
can be replaced
would lead to both deterministic coil selection and runtime. But so far,
no such threshold has been found.

Another possible extension is the generalization of coil selection to
coil weighting: If the origin of streak artifacts are high intensity regions
mostly outside of the \gls{FOV}, then simply weighing the coils appropriately
could lessen the severity of the artifacts while still including potentially
useful signal contained in these coils. Coil selection is then
simply a limiting case of coil weighting,
where the coils are assigned binary weights.

And finally, regularization of the reconstruction method
can be used to constrain the space of allowed solutions. So,
if a method of regularization which excludes images with strong
streak artifacts
from the solution space could be found, it would completely negate the need
for both coil selection
or coil weighting methods. So far however, no methods have been found
which reliably exclude streak artifacts from in-vivo acquisitions.

\FloatBarrier
\cleardoublepage
\chapter{Summary and Outlook}
\label{sec:Summary}
In this thesis, improvements for two data preprocessing steps in real-time
\gls{MRI} were described, evaluated, and discussed.

Starting from the problem arising from multiple receiver coils
in modern \gls{MRI}, the current \gls{PCA}-based compression is
introduced in \cref{sec:Compression}. From there, an optimized compression
algorithm is described and its adaptation to real-time \gls{MRI} is evaluated.
This  revealed difficulties with integrating the
optimized combination into the current real-time \gls{MRI} pipeline.
Furthermore, the improvement in image quality was judged as being too small
compared to these drawbacks to justify inclusion into
the real-time pipeline.

Based on two existing algorithms, this thesis developed and
evaluated
a new algorithm for coil selection in order to reduce
streak artifacts (\cref{sec:Selection}). The proposed
sinogram-based selection
provides significant improvement over existing algorithms for real-time
\gls{MRI}.
While still not
capable of removing all streak artifacts, it does provide mitigation
in a number of datasets.

Further work is necessary to identify the reasons for
this incomplete artifact removal. Another area warranting further
investigation is for acquisition modes other than anatomical:
In phase contrast flow \gls{MRI} \cite{PC-flow}, the main problem
is streak artifacts in the phase map. There, coil selection approaches
like the proposed sinogram-based selection do not provide a
benefit (see \cref{fig:sino_comparison_PC}). A reason for this is the
different streak origin: Here, susceptibility differences lead to
phase distortions which in turn lead to streak artifacts. In
\cref{fig:sino_comparison_PC}, the susceptibility difference is between
the venous blood and the surrounding brain tissue. Since these are phase
effects, magnitude-based coil selection methods are not suited for
these kinds of streak artifacts and alternatives need to be explored.

In general, sinogram-based selection leads to improved image quality
in a number of important acquisition schemes, including real-time \gls{MRI}
of the heart and head.
A \texttt{C++} implementation was therefore
added to \nlinvpp as part of the real-time pipeline and it is currently
used by default on all anatomical real-time \gls{MRI} scans
acquired in the institute.

\begin{figure}[tbp]
\centering
\foreach \index in {all,integr} {%
\begin{subfigure}{0.38\textwidth}
  \centering
 \includegraphics[width=0.95\textwidth]
 {figures/matlab_figs/T13480_liquor_flow_7spk/T13480_nlinvpp_\index_filter_crop_t3.png}
 \caption{\repr{\index} coils}
 \label{fig:sino_bad_liquor_\index}
\end{subfigure}\hspace{0.1\textwidth}
}\hspace{-0.1\textwidth}\vspace{1em}\\
\foreach \index in {all,integr} {%
\begin{subfigure}{0.38\textwidth}
  \centering
 \includegraphics[width=0.95\textwidth]
 {figures/matlab_figs/T13480_liquor_flow_7spk/T13480_nlinvpp_\index_filter_crop_t3_phase.png}
 \caption{\repr{\index} coils}
 \label{fig:sino_bad_liquor_phase_\index}
\end{subfigure}\hspace{0.1\textwidth}
}\hspace{-0.1\textwidth}
\caption[]{
Phase-contrast flow dataset \data{Flow1}, showing
a transversal slice through the brain.
\subref{fig:sino_bad_liquor_all} and \subref{fig:sino_bad_liquor_integr}
show the magnitude images with and without sinogram-based coil selection.
Neither image contains streak artifacts.
\subref{fig:sino_bad_liquor_phase_all} and
\subref{fig:sino_bad_liquor_phase_integr} show the corresponding
phase difference maps, which are proportional to flow velocity
perpendicular to the image plane.
\subref{fig:sino_bad_liquor_phase_all} contains streak
artifacts originating from a blood vessel in the
anterior region of the brain, the sagittal sinus,
which are not reduced by
coil selection \subref{fig:sino_bad_liquor_phase_integr}.
The streak ratios calculated by
sinogram-based selection can be found in \cref{fig:sino_method_Flow1}
in \cref{sec:figures}.
}
\label{fig:sino_comparison_PC}
\end{figure}

\appendix
\chapter{Additional Figures}
\label{sec:figures}
\newcommand*{\barDiagramWidth}{0.72\textwidth}

\noindent\begin{minipage}{\textwidth}
\captionsetup{type=figure}
\centering
  \includegraphics[width=\barDiagramWidth]
 {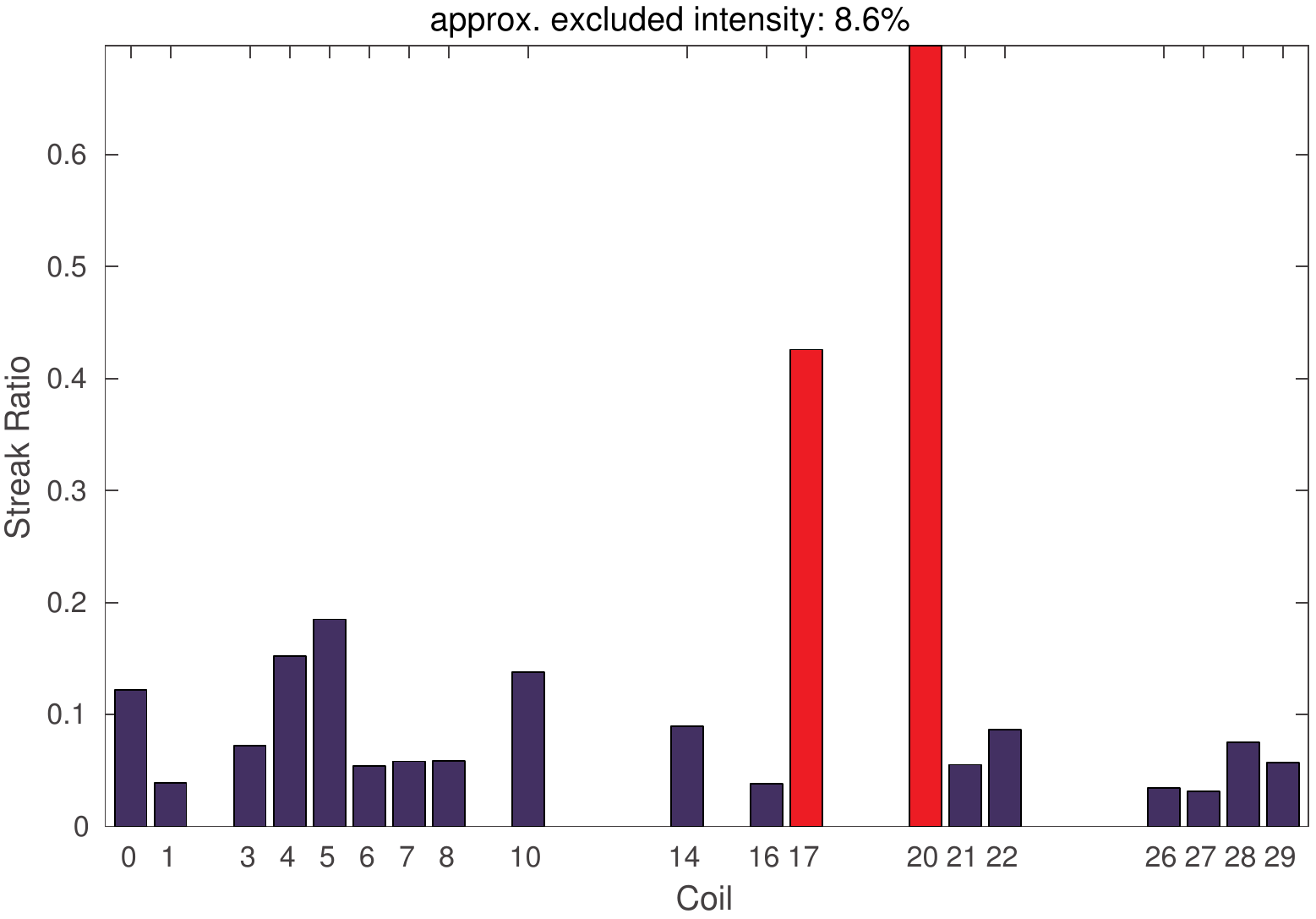}
\caption[]{
Result of the sinogram-based selection for dataset \data{Phan1}.
Coils in red are to be excluded.
}
\label{fig:sino_method_Phan1}
\end{minipage}
\vspace{1em}

\noindent\begin{minipage}{\textwidth}
\captionsetup{type=figure}
\centering
  \includegraphics[width=\barDiagramWidth]
 {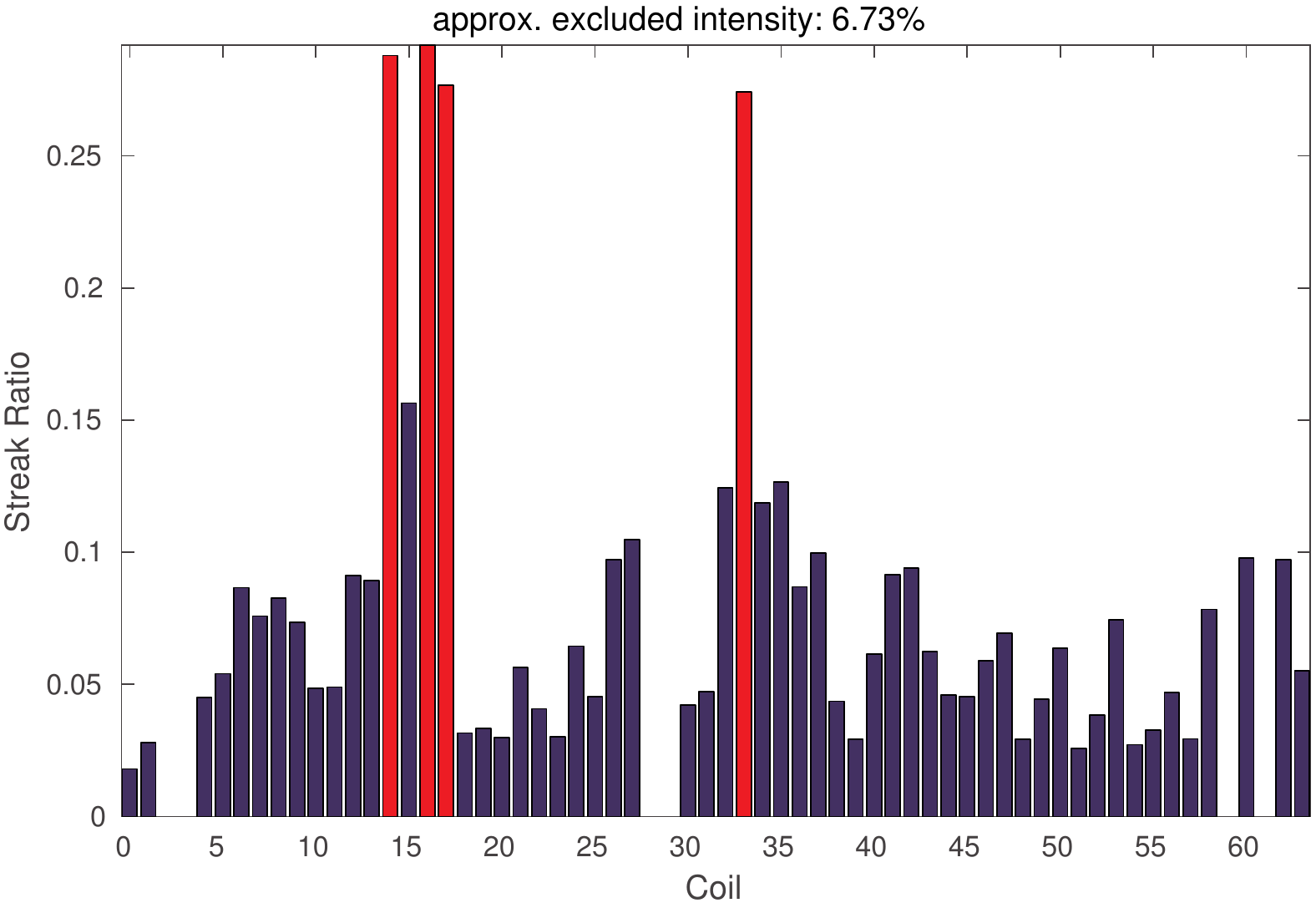}
\caption[]{
Result of the sinogram-based selection for dataset \data{Head1}.
Coils in red are to be excluded.
}
\label{fig:sino_method_Head1}
\end{minipage}
\newpage

\noindent\begin{minipage}{\textwidth}
\captionsetup{type=figure}
\centering
  \includegraphics[width=\barDiagramWidth]
 {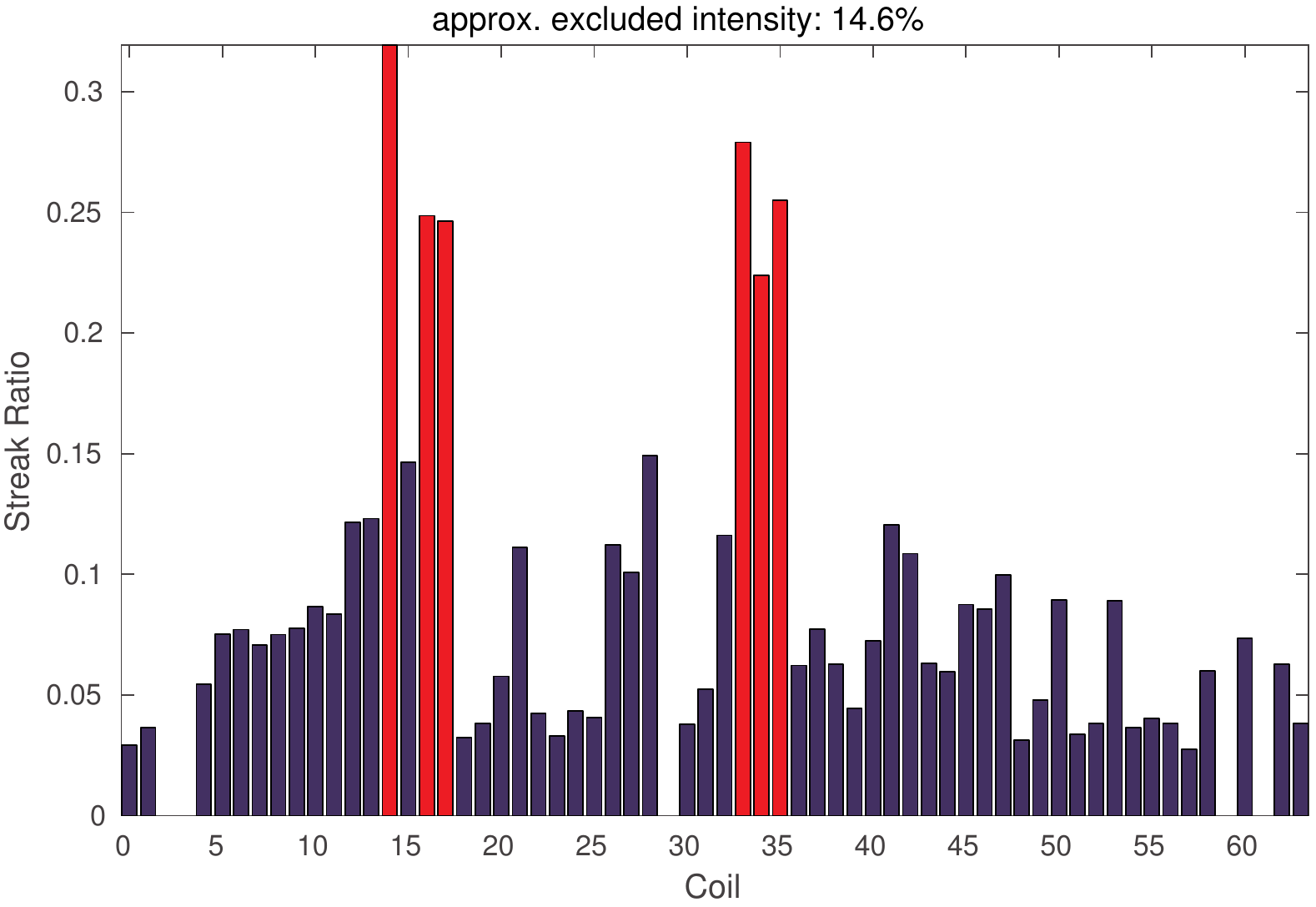}
\caption[]{
Result of the sinogram-based selection for dataset \data{Head2}.
Coils in red are to be excluded.
}
\label{fig:sino_method_Head2}
\end{minipage}
\vspace{1em}

\noindent\begin{minipage}{\textwidth}
\captionsetup{type=figure}
\centering
  \includegraphics[width=\barDiagramWidth]
 {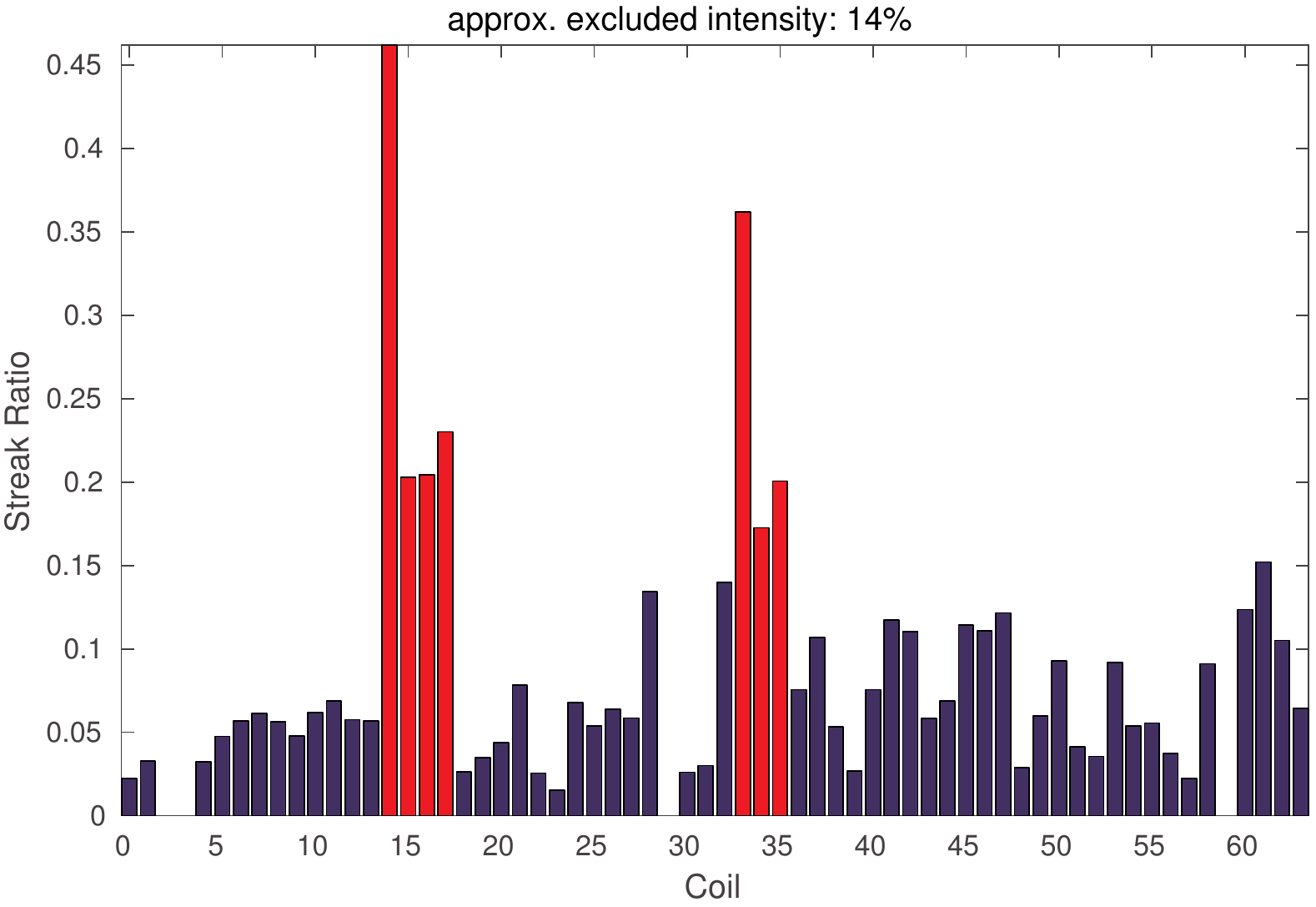}
\caption[]{
Result of the sinogram-based selection for dataset \data{Head4}.
Coils in red are to be excluded.
}
\label{fig:sino_method_Head4}
\end{minipage}
\newpage

\noindent\begin{minipage}{\textwidth}
\captionsetup{type=figure}
\centering
  \includegraphics[width=\barDiagramWidth]
 {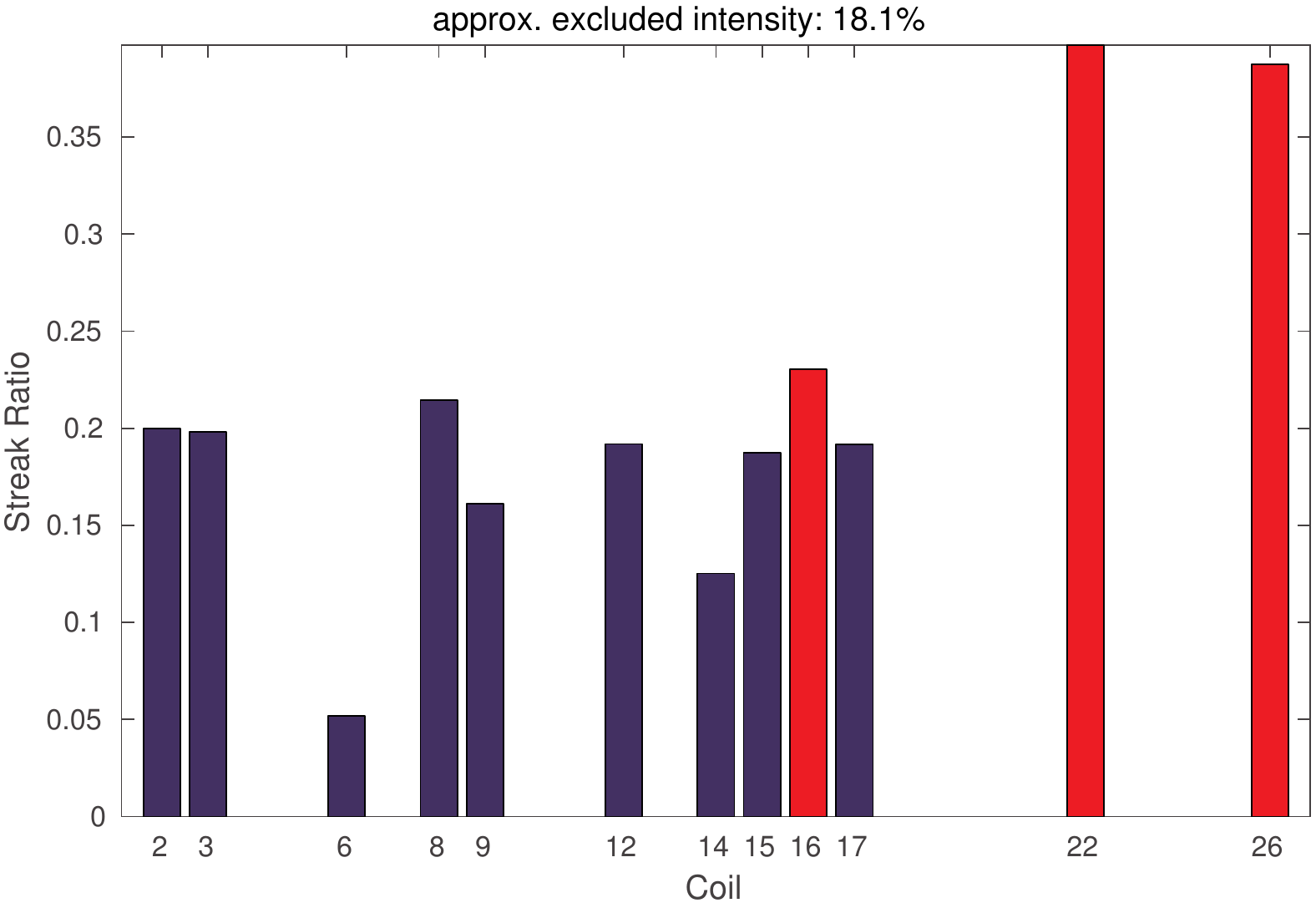}
\caption[]{
Result of the sinogram-based selection for dataset \data{Flow1}.
Coils in red are to be excluded.
}
\label{fig:sino_method_Flow1}
\end{minipage}


\raggedbottom
\FloatBarrier
\clearpage
\chapter{Sequence Parameters}
\label{sec:sequences}

\begin{table}[tbh]
\centering
 \caption{
 Descriptions of sequences used in this thesis.
 }
 \label{tab:descriptions}
{\def\arraystretch{1.5}
\begin{tabular}{lp{0.7\textwidth}}
\toprule
  \data{Phan1}  & Static commercial water phantom.\\
  \data{Phan2}  & Perfusion phantom during active flow.\\
  \data{Head1}  & Midsagittal slice of the head during speaking.
                    Human male with no known illnesses, 24 years old.\\
  \data{Head2}  & Midsagittal slice of the head during speaking.
                    Human male with no known illnesses, 28 years old.\\
  \data{Head3}  & Midsagittal slice of the head during speaking.
                    Human male with no known illnesses, 26 years old.\\
  \data{Head4}  & Midsagittal slice of the head while
                    playing a brass instrument.
                    Human male with no known illnesses, 48 years old.\\
  \data{Head5}  & Midsagittal slice of the head while
                    playing a brass instrument.
                    Human female with no known illnesses, 26 years old.\\
  \data{Heart1} & Short-axis view of the heart.
                  Human male with no known illnesses, 25 years old.\\
  \data{Heart2} & Short-axis view of the heart.
                  Human male with no known illnesses, 20 years old.\\
  \data{Heart3} & Short-axis view of the heart.
                  Human male with no known illnesses, 36 years old.\\
  \data{Flow1}  & Phase contrast flow acquisition of a transversal
                  slice through the brain.
                  Human male with no known illnesses, 33 years old.\\
\bottomrule
\end{tabular}
}
\end{table}
\raggedbottom

\newcolumntype{C}{>{\collectcell\usermacro}l<{\endcollectcell}}
\newcommand{\usermacro}[1]{\ifbool{dirk}{#1}{\hspace*{-2\tabcolsep}}}

\newbool{dirk}
\setbool{dirk}{true}

\begin{landscape}
\begin{center}
\begin{longtable}{l
                  S[table-format=1.2]
                  S[table-format=1.2]
                  S[table-format=2]
                  c
                  S[table-format=2]
                  cc
                  S[table-format=3]
                  S[table-format=4]
                  C
                  r
                  }
\caption{
Overview over sequences parameters.
}
\label{tab:sequences}\\
\toprule
 \multirow{2}{*}{\textbf{Label}} & {$\bm{T_R}$} & {$\bm{T_E}$} &
  $\bm{n_\text{spokes}}$ & $\bm{n_\text{turns}}$ & $\bm{\alpha}$ &
  \textbf{\glsunset{FOV}\gls{FOV}} & \textbf{Resolution} &
  {\textbf{Frame rate}} & {\textbf{Bandwidth}} &
  \multirow{2}{*}{\textbf{TNumber}} & \hspace*{-2\tabcolsep} \\
 & {[\si{\ms}]} & {[\si{\ms}]} & [{1}] & [1] & [\si{\degree}] &
  [\si{\square\mm}] & [\si{\cubic\mm}] & {[\si{\per\second}]} &
  {[\si{\hertz\per\pixel}]} & & \hspace*{-2\tabcolsep} \\
\midrule
\endfirsthead
\multicolumn{9}{l}%
{\tablename\ \thetable\ -- \textit{Continued from previous page}} \\
\midrule
 \multirow{2}{*}{\textbf{Label}} & {$\bm{T_R}$} & {$\bm{T_E}$} &
  $\bm{n_\text{spokes}}$ & $\bm{n_\text{turns}}$ & $\bm{\alpha}$ &
  \textbf{\glsunset{FOV}\gls{FOV}} & \textbf{Resolution} &
  {\textbf{Frame rate}} & {\textbf{Bandwidth}} &
  \multirow{2}{*}{\textbf{TNumber}} & \hspace*{-2\tabcolsep} \\
 & {[\si{\ms}]} & {[\si{\ms}]} & [{1}] & [1] & [\si{\degree}] &
  [\si{\square\mm}] & [\si{\cubic\mm}] & {[\si{\per\second}]} &
  {[\si{\hertz\per\pixel}]} & & \hspace*{-2\tabcolsep} \\
\midrule
\endhead
\multicolumn{9}{r}{\textit{Continued on next page}} \\
\endfoot
\bottomrule
\endlastfoot
\data{Phan1} 
    & 1.63
    & 0.95
    & 7
    & 7
    & 10
    & $256{\times}256$
    & $1.6{\times}1.6{\times}6$
    & 30
    & 1955
    & T3506
    &\hspace*{-2\tabcolsep}\\
\data{Phan2} 
    & 2.32
    & 1.58
    & 17
    & 5
    & 8
    & $192{\times}192$
    & $1.2{\times}1.2{\times}6$
    & 25
    & 1565
    & T11500
    &\hspace*{-2\tabcolsep}\\
\data{Head1} 
    & 1.96
    & 1.28
    & 17
    & 5
    & 5
    & $192{\times}192$
    & $1.5{\times}1.5{\times}10$
    & 30
    & 1860
    & T6954
    &\hspace*{-2\tabcolsep}\\
\data{Head2} 
    & 2.02
    & 1.28
    & 9
    & 7
    & 5
    & $192{\times}192$
    & $1.4{\times}1.4{\times}8$
    & 55
    & 1600
    & T16936
    &\hspace*{-2\tabcolsep}\\
\data{Head3}
    & 2.0
    & 1.28
    & 5
    & 9
    & 5
    & $192{\times}192$
    & $1.4{\times}1.4{\times}8$
    & 100
    & 1670
    & T17086
    &\hspace*{-2\tabcolsep}\\
\data{Head4} 
    & 1.96
    & 1.23
    & 17
    & 5
    & 5
    & $192{\times}192$
    & $1.5{\times}1.5{\times}10$
    & 30
    & 1630
    & T25921
    &\hspace*{-2\tabcolsep}\\
\data{Head5} 
    & 2.02
    & 1.28
    & 9
    & 7
    & 5
    & $192{\times}192$
    & $1.4{\times}1.4{\times}8$
    & 55
    & 1600
    & T17037
    &\hspace*{-2\tabcolsep}\\
\data{Heart1} 
    & 1.96
    & 1.22
    & 17
    & 5
    & 8
    & $256{\times}256$
    & $1.6{\times}1.6{\times}6$
    & 30
    & 1565
    & T18907
    &\hspace*{-2\tabcolsep}\\
\data{Heart2} 
    & 2.22
    & 1.28
    & 5
    & 5
    & 16
    & $256{\times}256$
    & $1.6{\times}1.6{\times}6$
    & 30
    & 1040
    & T10366
    &\hspace*{-2\tabcolsep}\\
\data{Heart3} 
    & 2.26
    & 1.47
    & 27
    & 5
    & 4
    & $256{\times}256$
    & $1.0{\times}1.0{\times}6$
    & 16
    & 1395
    & T16228
    &\hspace*{-2\tabcolsep}\\
\data{Flow1} 
    & 5.19
    & 4.36
    & 13
    & 5
    & 10
    & $192{\times}192$
    & $1.2{\times}1.2{\times}5$
    & 7
    & 1250
    & T13480
    &\hspace*{-2\tabcolsep}\\
\end{longtable}
\end{center}
\end{landscape}
\cleardoublepage
\printbibliography[heading=bibintoc]

\chapter*{Acknowledgements}
\iffalse
Ich glaube ich werde wohl irgendwem wohl irgendwofür danken....
\else
First, I would like to thank Prof. Dr. Jens Frahm for the
excellent supervision during this thesis and for giving me the
opportunity to pursue my research in such a distinguished environment.

I would also like to thank Prof. Dr. Tim Salditt for reviewing this
thesis as second referee.

Furthermore, I would like to thank my colleagues at the BiomedNMR,
for providing an engaging and thoroughly cooperative atmosphere, and
for the many helpful discussions during my thesis.
Here, I would particularly like to thank Volkert Roeloffs, Dr. Dirk Voit,
Zhengguo Tan, Jakob Klosowski,
Dr. Oleksandr Kalentev, and Sebastian Schätz.

I am deeply indebted to Christina Bömer and Marie Zeiß for their
unwavering emotional support and encouragement.

I would like to thank Prof. Dr. Thomas Pruschke for providng
the \LaTeX{} template used to write this thesis.
\fi

\Declaration
\end{document}